\documentclass[11pt]{article}
\usepackage{amsthm,amsmath,latexsym,amssymb,amsfonts,amsbsy}
\usepackage{graphicx,lscape,fancyhdr,color,array,stmaryrd,euscript,wrapfig}
\usepackage{cite}
\newcommand{\bep}{\begin{picture}}
\newcommand{\eep}{\end{picture}}
\newcounter{YoungHeight}\newcounter{YoungWidth}
\newcounter{Mul1}\newcounter{Mul2}
\newcounter{A1}\newcounter{A2}
\newcounter{T0}\newcounter{T1}

\newlength{\txtHShift}

\newlength{\txtWidth}

\newcommand{\HalfLength}[2]{\setcounter{Mul1}{#1}\setcounter{Mul2}{#1}\addtocounter{Mul1}{\value{Mul2}}\addtocounter{Mul1}{\value{Mul2}}%
\addtocounter{Mul1}{\value{Mul2}}\addtocounter{Mul1}{\value{Mul2}}\setcounter{#2}{\value{Mul1}}}

\newcommand{\Add}[3]{\setcounter{#1}{#2}\addtocounter{#1}{#3}}

\newcommand{\Length}[1]{#10}
\newcommand{\YoungScale}{}

\newcommand{\shiftedText}[2]{{\hspace{#1}#2}}
\newcommand{\calcHShift}[1]{\settowidth{\txtWidth}{#1}\setlength{\txtHShift}{-0.5\txtWidth}}

\newcommand{\TextTop}[3]{{\calcHShift{#1}\HalfLength{#2}{T0}\Add{T1}{\Length{#3}}{-9}\put(\value{T0},\value{T1}){\shiftedText{\txtHShift}{#1}}}}

\newcommand{\RectT}[3]{\bep(\Length{#1},\Length{#2})\put(0,0){\line(1,0){\Length{#1}}}\put(0,0){\line(0,1){\Length{#2}}}%
\put(\Length{#1},\Length{#2}){\line(-1,0){\Length{#1}}}\put(\Length{#1},\Length{#2}){\line(0,-1){\Length{#2}}}#3{#1}{#2}\eep}

\newcommand{\RectARow}[2]{{\bep(\Length{#1},10)\put(0,0){\RectT{#1}{1}{\TextTop{#2}}}\eep}}

\newcommand{\RectBRow}[4]{{\bep(\Length{#1},20)\put(0,0){\RectT{#2}{1}{\TextTop{#4}}}%
\put(0,10){\RectT{#1}{1}{\TextTop{#3}}}\eep}}

\newcommand{\BlockApar}[2]{\parbox{\Length{#1}pt}{\YoungScale\bep(\Length{#1},\Length{#2}){\Add{A1}{#1}{1}\Add{A2}{#2}{1}}%
\multiput(0,0)(10,0){\value{A1}}{\line(0,1){\Length{#2}}}\multiput(0,0)(0,10){\value{A2}}{\line(1,0){\Length{#1}}}%
\setcounter{YoungHeight}{\Length{#2}}\setcounter{YoungWidth}{\Length{#1}}\eep}}

\newcommand{\YoungpAAA}{\BlockApar{1}{3}}
\newcommand{\YoungpAAAA}{\BlockApar{1}{4}}
\newcommand{\be}{\begin{equation}}
\newcommand{\ee}{\end{equation}}
\newcommand{\pl}{\partial}

\newcommand{\boldpic}[1]{{\linethickness{0.4mm}#1}}
\renewcommand{\theequation}{\arabic{section}.\arabic{equation}}

\newcommand{\fud}[2]{{^{#1}_{\phantom{#1}#2}}}
\newcommand{\fdu}[2]{{_{#1}^{\phantom{#1}#2}}}

\newcommand{\fudu}[3]{{^{#1\phantom{#2}#3}_{\phantom{#1}#2}}}

\definecolor{rougef}{rgb}{0.56,0,0}
\definecolor{vertf}{rgb}{0,0.5,0}
\definecolor{bleuf}{rgb}{0,0,0.8}

\def\sideremark#1{\ifvmode\leavevmode\fi\vadjust{\vbox to0pt{\vss
 \hbox to 0pt{\hskip\hsize\hskip1em
 \vbox{\hsize2cm\tiny\raggedright\pretolerance10000
 \noindent #1\hfill}\hss}\vbox to8pt{\vfil}\vss}}}%

\begin{document}
\begin{titlepage}

\begin{flushright}
\vspace{1mm}
AEI-2013-212
\end{flushright}

\vspace{1cm}
\begin{center}
{\bf \Large On the uniqueness of higher-spin symmetries in AdS and CFT}
\vspace{2cm}

\textsc{N. Boulanger\footnote{Research Associate of the Fund
for Scientific Research-FNRS (Belgium);
nicolas.boulanger@umons.ac.be}, D.
Ponomarev\footnote{dmitri.ponomarev@umons.ac.be},
E. Skvortsov\footnote{skvortsov@lpi.ru}
and M. Taronna\footnote{massimo.taronna@aei.mpg.de}}

\vspace{2cm}

{\em${}^{1,2}$  Service de M\'ecanique et Gravitation, UMONS,
20 Place du Parc, 7000 Mons, Belgium}

{\em${}^{3,4}$  Albert Einstein Institute, Golm, Germany, D-14476, Am M\"{u}hlenberg 1}

{\em${}^{3}$  Lebedev Institute of Physics, Moscow, Russia, 119991, Leninsky pr-t, 53}

\end{center}

\vspace{0.5cm}
\begin{abstract}

We study the uniqueness of higher-spin algebras which are at the core of higher-spin theories in AdS and of CFTs with exact higher-spin symmetry, {\emph{i.e.}} conserved tensors of rank
greater than two.
The Jacobi identity for the gauge algebra is the simplest consistency test that appears at the quartic order for a gauge theory. Similarly, the algebra of charges in a CFT must also obey the Jacobi identity. These algebras are essentially the same. Solving the Jacobi identity under some simplifying assumptions spelled out, we obtain that the
Eastwood-Vasiliev algebra is the unique solution for $d=4\,$ and $d\geqslant 7$. In $5d$ there is a one-parameter family of algebras that was known before.
In particular, we show that the introduction of a single higher-spin
gauge field/current automatically requires the infinite tower of
higher-spin gauge fields/currents.
The result implies that from all the admissible non-Abelian cubic vertices in $AdS_d$, that
have been recently classified for totally symmetric
higher-spin gauge fields, only one vertex
can pass the Jacobi consistency test.
This cubic vertex is associated with a gauge deformation
that is the germ of the Eastwood-Vasiliev's higher-spin algebra.
\end{abstract}

\end{titlepage}

\numberwithin{equation}{section}

\section{Introduction}
\label{intro}
There are two questions about higher-spin symmetries that are closely
related via the AdS/CFT correspondence. How many higher-spin theories
in $AdS_d$ there exist? How many $CFT_{d-1}$'s with conserved
currents of rank higher than two there exist?
In $AdS_d$ one starts with
a quadratic Lagrangian (linear equations of motion) and tries
to add cubic,
quartic vertices etc. deforming the gauge transformations in such a way
as to make the Lagrangian (equations of motion) gauge invariant.
In a $CFT_{d-1}$ with a stress-tensor and higher-spin conserved
currents one constructs the corresponding charges, studies their action
on operators and attempts to solve Ward identities, \cite{Maldacena:2011jn}.
On both sides of the AdS/CFT correspondence the two procedures lead
to severe restrictions on the spectrum of fields in $AdS_d$ and on the
spectrum of operators in the $CFT_{d-1}$.

Answering both questions boils down to the purely algebraic problem of classifying higher-spin algebras. A higher-spin (HS) algebra can be viewed either as generating rigid symmetries of HS theories in $AdS_d$ or as the algebra of charges in $CFT_{d-1}$, and as such it is related to the fusion algebra of conserved HS currents.

While the question of uniqueness of higher-spin algebras was addressed long ago in $AdS_4$ by Fradkin
and Vasiliev \cite{Fradkin:1986ka}, an equivalent problem for $CFT_3$ was solved only recently by
Maldacena and Zhiboedov \cite{Maldacena:2011jn}. In the present paper we study the problem of
classifying higher-spin algebras in $AdS_d$ with $d>3$ and solve it under certain assumptions in $d=4$ and $d\geqslant7\,$, with the conclusion that the higher-spin algebra that is relevant for totally-symmetric higher-spin fields in $AdS_d$, and for $CFT_{d-1}$'s with exactly conserved totally-symmetric higher-spin tensors, is unique. $AdS_5$ is a special case where one can have a one-parameter family of higher-spin algebras found in \cite{Boulanger:2011se} and, from a different perspective, in \cite{Fernando:2009fq}. We do not have any exhaustive classification for this case. The $d=6$ case is not covered too due to the absence of an effective formalism that implements certain Schouten-like identities. Basically, our proof amounts to identifying the possible structures that can contribute to the commutator of generators reducing the problem to a well-known problem in deformation quantization.

Let us now outline the precise relation between AdS and CFT setups, \cite{Didenko:2012vh}, (see also \cite{Taronna:2012gb} for some related discussions and ideas). In $AdS_d$ we have a theory whose spectrum contains the graviton and at least one higher-spin gauge field that at the free level can be described by a rank-$s$ totally-symmetric Fronsdal field \cite{Fronsdal:1978rb},
\be\label{frfield}
\delta \phi_{\mu_1\ldots\mu_s}=\nabla_{\mu_1} \xi_{\mu_2\ldots\mu_s}+\mbox{permutations}\,.
\ee
The boundary values of the Fronsdal field at the conformal infinity of AdS are gauge fields themselves, the Fradkin-Tseytlin conformal higher-spin fields \cite{Fradkin:1985am}, $\mu=\{a,z\}$:
\be
\left.\phi_{a_1\ldots a_s}\rule{0pt}{12pt}\right|_{z\rightarrow0}=z^\Delta \bar{\phi}_{a_1\ldots a_s}+\ldots\,, \quad\qquad\qquad  \delta \bar{\phi}_{a_1\ldots a_s}=
(\pl_{a_1} \bar{\xi}_{a_2\ldots a_s} +\mbox{permutations}) -
(\mbox{traces})\,,
\ee
that naturally couple  to conserved tensors
\be \label{FRT}
\Delta S=\int d^{d-1}x\, j_{a_1\ldots a_s} \bar{\phi}^{a_1\ldots a_s}\,,\qquad\qquad\qquad \pl^m j_{ma_2\ldots a_s}=0\,.
\ee
According to the standard AdS/CFT dictionary \cite{Maldacena:1997re, Gubser:1998bc, Witten:1998qj},
a conserved tensor is an operator that is dual to a bulk higher-spin gauge field. Therefore, the CFT dual
theory of a higher-spin gauge theory must have conserved tensors in the spectrum in addition to the
stress-tensor. For instance, the CFT stress-tensor $j_{a_1a_2}$ can be used to construct the full set of
charges corresponding to the conformal algebra $so(d-1,2)$. This is done by contracting it with
conformal Killing vectors to get conserved currents. Analogously, given a higher-spin conserved tensor
$j_{a_1\ldots a_s}$, one can construct various conserved currents
\begin{align}
\pl^m j^s_m&=0\,, && j^s_m=j_{m a_2\ldots a_s} K^{a_2\ldots a_s}\,, && \mbox{trace free part of }(\pl^{a_1}K^{a_2\ldots a_s}+\mbox{perm})=0\,,
\end{align}
by contracting it with a conformal Killing tensor $K^{a_2\ldots a_s}$. The space of conformal Killing tensors of rank-$(s-1)$ is known \cite{Eastwood:2002su} to form an irreducible representation of $so(d-1,2)$ transforming as a tensor with the symmetry of a two-row rectangular Young diagram of length-$(s-1)$\,:
\begin{align}
&K_{a_2\ldots a_s}(x;\epsilon)=K_{a_2\ldots a_s}^{A(s-1),B(s-1)}(x)\,\epsilon_{A(s-1),B(s-1)}\,, && {A(s-1),B(s-1)}:\parbox{42pt}{\boldpic{\RectBRow{4}{4}{$s-1$}{}}}\,,
\end{align}
where\footnote{$A(s-1)\equiv A_1...A_{s-1}$ denotes the group of indices in which a tensor is symmetric. } $K_{a_2\ldots a_s}^{A(s-1),B(s-1)}(x)$ are fixed polynomial intertwining functions, and $\epsilon_{A(s-1),B(s-1)}$ are parameters. In particular for $s=2$, {\emph{i.e.}} stress-tensor, we have $\epsilon^{A,B}=-\epsilon^{B,A}$, whose number of components coincides with the dimension of $so(d-1,2)$, {\emph{i.e.}} the number of independent conformal Killing vectors.

The full set of charges associated with higher-spin conserved tensors $j_{a_1\ldots a_s}$ is thus labeled by conformal Killing tensors and can be constructed in a standard way by defining $(d-2)$-forms that are Hodge duals to the conserved currents $j^s_m$, which depend on the parameters $\epsilon_{A(s-1),B(s-1)}$\,:
\begin{align}
&Q(\epsilon)=Q^{A(s-1),B(s-1)}\epsilon_{A(s-1),B(s-1)}=\oint_{\Sigma_{d-2}} \Omega(\epsilon)\,, && \Omega(\epsilon)=\star j^s_m  (\epsilon)\, dx^m\,.
\end{align}

Having defined higher-spin charges one can study their action on various operators in the CFT, in particular on the conserved tensors themselves, and try to read off the constraints on the operator spectrum implied by the Ward identities. In the case of $CFT_3$ this was done in \cite{Maldacena:2011jn} with the result that the presence of at least one higher-spin conserved tensor goes in hand with the presence of infinitely many of them, whose spins range from zero to infinity, while their correlation functions admit a free field realization (either by a free boson or by a free fermion). In other words, any $CFT_3$ with exact higher-spin symmetry is essentially a free theory. This does not imply the triviality of the $AdS$-dual as the HS algebra is deformed at the interaction level and a theory may have interesting interacting CFT-duals for different choices of boundary conditions, \cite{Sezgin:2002rt,Klebanov:2002ja}.

From a slightly different perspective the Maldacena-Zhiboedov work amounts to a classification of higher-spin algebras, {\emph{i.e.}} the algebras that can be realized by the charges $Q(\epsilon)$ above. The starting point is to assume that there are at least two conserved tensors: the stress-tensor, $s=2$, and some other conserved higher-spin tensor, $s>2$. Therefore, we have at least two charges, $Q_2=Q(\epsilon^{A,B})$ and $Q_s=Q(\epsilon^{A(s-1),B(s-1)})$, and we can study the algebra they form by investigating the r.h.s. of $[Q_2, Q_s]=...$ and $[Q_s,Q_s]=...\,$.
By the CFT axioms some of the structure constants must be non-vanishing, e.g. $[Q_2, Q_s]=Q_s+...$ and $[Q_s,Q_s]=Q_2+...$, which via the Ward/Jacobi identities implies that some other structure constants must be non-zero as well. In particular, one concludes that it is not possible for Ward identities to be satisfied unless there are higher-spin charges of all spins (at least even), which extends the Maldacena-Zhiboedov result to higher dimensions. The HS algebra of charges $Q(\epsilon)$ is closely related to the fusion algebra $j_{s_1}\times j_{s_2}=\sum_s j_s+...$ of conserved tensors.

Back to $AdS_d$ and its conformal boundary $M_{d-1}=\pl AdS_d$, if one wishes to make higher-spin symmetries manifest one can couple to gauge fields the full multiplet of conserved currents associated with a conserved tensor, which is parameterized by $\epsilon_{A(s-1),B(s-1)}$ via conformal Killing tensors:
\begin{align}
&\Delta S=\int_{M_{d-1}} \Omega_{A(s-1),B(s-1)}\wedge \bar{W}^{A(s-1),B(s-1)}\,, && \delta \bar{W}^{A(s-1),B(s-1)}=D\xi^{A(s-1),B(s-1)}\,.
\end{align}
Here, $\bar{W}$ has to be a one-form taking values in the same representation of the conformal algebra and it is a gauge field again since $\Omega$ is a closed form, while $D$ is the $so(d-1,2)$-covariant derivative on $M_{d-1}=\pl AdS_d$. More in detail, it is a conformal higher-spin connection studied in \cite{Vasiliev:2009ck} and it contains the Fradkin-Tseytlin fields of eq.~\eqref{FRT}, which are boundary values of the Fronsdal fields. Viewing $\bar{W}^{A(s-1),B(s-1)}$ as a boundary value, we arrive at the conclusion that a natural way to formulate the AdS dual theory is to make use of one-forms
\begin{align} \label{vconn}
W^{A(s-1),B(s-1)}\,, && \delta W^{A(s-1),B(s-1)}=D\xi^{A(s-1),B(s-1)}\,.
\end{align}
These are the higher-spin connections introduced by Vasiliev in \cite{Vasiliev:2001wa} and applied later
to the construction of a higher-spin theory in \cite{Vasiliev:2003ev}.
Now $D$ is the $so(d-1,2)$-covariant derivative in $AdS_d$.
The Fronsdal field \eqref{frfield} is a particular component of \eqref{vconn}.

Analogously to charges $Q^{A(s-1),B(s-1)}$, forming an algebra as a consequence of the CFT axioms, the Vasiliev connections $W^{A(s-1),B(s-1)}$ are gauge fields of a higher-spin algebra \cite{Eastwood:2002su,Vasiliev:2003ev}, which is at the core of the higher-spin theory of \cite{Vasiliev:2003ev}. The question we would like to address in the present note is whether this algebra is unique or not. This question turns out to be closely related to non-Abelian cubic vertices of HS fields.

Recently \cite{Vasilev:2011xf, Boulanger:2012dx} all the possible
non-Abelian cubic couplings between totally-symmetric higher-spin (including spin-$2$)
gauge fields in $AdS_d\,$, with $d\geqslant 4\,$ have been explicitly built and classified (see also \cite{Joung:2011ww,Joung:2012rv,Joung:2012hz,JoungTaronnaToAppear} for the corresponding classification, in the metric-like formalism, of (non-)Abelian vertices together with the analysis of non-trivial deformations of gauge transformations)\footnote{Results on cubic vertices in Minkowski space were obtained in \cite{Metsaev:1993ap, Metsaev:1993mj,Metsaev:2005ar,Manvelyan:2010jr,Sagnotti:2010at,Bekaert:2010hp}. See also references in \cite{Bekaert:2010hw}.}. Some simplifications resulted from using the manifestly $AdS_d$-covariant frame-like formalism of \cite{Vasiliev:2001wa}.
By non-Abelian cubic vertices we mean those which non-trivially
deform the Abelian gauge algebra of the free theory. They are obviously defined up to the addition of Abelian vertices.
Actually, a precise definition can be given within the BRST cohomological language and can be found in
\cite{Barnich:1993vg}.
The main result of \cite{Vasilev:2011xf, Boulanger:2012dx} is that given three totally-symmetric gauge
fields with spins $s$, $s'$ and $s''$, the number of inequivalent non-Abelian vertices
is given by the tensor product multiplicity
\be \label{tworowmul}\mbox{Number of singlets in}\left[\rule{0pt}{22pt}\parbox{42pt}{\boldpic{\RectBRow{4}{4}{$s-1$}{}}}\bigotimes
\parbox{50pt}{\boldpic{\RectBRow{5}{5}{$s'-1$}{}}}\bigotimes
\parbox{60pt}{\boldpic{\RectBRow{6}{6}{$s''-1$}{}}}\right]\ ,  \ee
{\emph{i.e.}} by all the possible independent ways to contract two
$so(d-1,2)$ tensors with the Young symmetry specified above in order to form another $so(d-1,2)$ tensor
of the same Young symmetry type. This follows directly from the Fradkin-Vasiliev construction
\cite{Fradkin:1987ks,Fradkin:1986qy,Vasiliev:2011xf}, where the deformations of the linearized
field-strengths are of the Yang-Mills type and are constructed by adding all possible terms bilinear in the HS gauge connections \eqref{vconn}
\begin{align}\label{fvdeform}
R^{s}&=D W^{s} + W^{s'}\diamond W^{s''}\,,\\
&W^{s'}\diamond W^{s''}=\mbox{projection onto \parbox{42pt}{\boldpic{\RectBRow{4}{4}{$s-1$}{}}} of }\left(W^{A(s'-1),B(s'-1)} \bigotimes W^{A(s''-1),B(s''-1)}\right)\,.\nonumber
\end{align}
Analogously, the above statement can be interpreted by saying that the number of independent non-Abelian cubic vertices is in one to one correspondence with the number of non-trivial global symmetry structure constants that can be built starting from the corresponding Killing tensors associated with massless fields, see e.g.
\cite{Bekaert:2005ka,Bekaert:2006us,Bengtsson:2008mw,Sagnotti:2010at,Joung:2011ww} and further comments below.
A particular way of contracting indices in (\ref{tworowmul}) or \eqref{fvdeform} is given by the Eastwood-Vasiliev
algebra \cite{Eastwood:2002su,Vasiliev:2003ev}, which is the {\it unique associative algebra} obtained by quotienting the
universal enveloping algebra of $so(d-1,2)$ so as to leave generators that are irreducible $so(d-1,2)$-tensors with the symmetry of \eqref{fvdeform}.
One can recover the HS Lie algebra by taking the commutator as Lie bracket.
The result is Vasiliev's simplest higher-spin Lie algebra \cite{Vasiliev:2003ev}.

The non-Abelian deformations that were classified in \cite{Boulanger:2012dx} are
uniquely associated with certain $\diamond$-products, \eqref{fvdeform}, {\emph{i.e.}}
they are given by linearly independent elements of the tensor product of the specific $so(d-1,2)$-modules. That the tensor product is not multiplicity-free results in a number of non-Abelian cubic couplings for a fixed triple of spins. Most of the $\diamond$-products define non-associative algebras. Whether a given $\diamond$-product
gives rise to any associative or Lie structure is irrelevant for the construction of cubic vertices.
The algebraic properties of the $\diamond$-products become important at the next order in perturbation,
{\emph{viz.}} quartic level.
Indeed, it is easy to see that the non-Abelian cubic vertices
which have a chance to be consistently completed by quartic
ones are those for which the corresponding
gauge algebra satisfies the Jacobi identity.

In this paper, we solve the Jacobi identity for all
the possible $\diamond$-products obtained in \cite{Boulanger:2012dx}
and show that the only solution, under assumptions that
we spell out below, is the Eastwood-Vasiliev higher-spin
algebra \cite{Eastwood:2002su, Vasiliev:2003ev}.
Our assumptions are as follows:
\begin{itemize}
\item Since the Jacobi identity arises at the quartic level, it is necessary to require the theory to pass the cubic test first. This amounts in the frame-like formalism to the Fradkin-Vasiliev procedure \cite{Fradkin:1986qy,Fradkin:1987ks}.
This condition can be translated in mathematical terms saying that the resulting gauge algebras
at that order have an invariant norm or equivalently that there exists a non-degenerate Killing metric, so that the structure constants can be made totally anti-symmetric. On the CFT side this is equivalent to the fact that two-point correlation functions define a non-degenerate norm, whose existence implies certain reflection properties for the structure constants of the hypothetical algebra, see \cite{Maldacena:2011jn};

\item The algebra $so(d-1,2)$ is a Lie subalgebra of
the higher-spin algebra, and the higher-spin
generators, whenever present, transform as tensors under the adjoint action
of $so(d-1,2)\,$.

For a higher-spin theory in AdS this implies the presence of a graviton in the spectrum. It also implies
that higher-spin fields, whenever present in the spectrum, interact minimally with gravity, {\emph{i.e.}} via the Lorentz-covariantization of derivatives.
On the CFT side, we have a stress-tensor.
In case there exists also a conserved higher-spin tensor,  then it
transforms canonically under the action of the generators of the conformal algebra constructed from the stress-tensor itself;

\item There must be at least one higher-spin generator in the spectrum.
The higher-spin generators do not carry any additional flavor indices, i.e. the multiplicity of a generator is either zero or one.
\end{itemize}

All together we are looking for a Lie algebra that
has at least two generators --- the generator $T_{AB}$ of the conformal/AdS algebra $so(d-1,2)$, which obeys
\be
[T_{AB},T_{CD}]=T_{AD}\eta_{BC}+\mbox{permutations}\,,
\label{AdSalgebra}
\ee
and a higher-spin generator $T_{A(s-1),B(s-1)}$ that is an irreducible $so(d-1,2)$-tensor with the symmetry of \parbox{42pt}{\boldpic{\RectBRow{4}{4}{$s-1$}{}}} obeying
\be
\label{transformastensors}
[T_{AB}, T_{C(s-1),D(s-1)}]= T_{AC(s-2),D(s-1)}\eta_{BC}+\mbox{permutations}\,,
\ee
{\emph{i.e.}} gravity interacts minimally with a higher-spin field or, in the CFT,
the higher-spin conserved tensor transforms properly under the conformal algebra transformations.
\footnote{\label{footnote7}Actually, one could weaken the assumptions
\eqref{AdSalgebra} and \eqref{transformastensors} by using the
well-know results about the uniqueness of the spin-2
non-Abelian self-coupling and the uniqueness of the gravitational
coupling of totally symmetric higher-spin gauge fields
in $AdS_{d\geqslant 4}$ background, so that \emph{stricto senso} only \eqref{transformastensors} is necessary. The existence of the subalgebra \eqref{AdSalgebra} follows from it.}

The Fradkin-Vasiliev condition, or the anti-symmetry of the structure constants, or the fact that the stress-tensor must appear in the OPE of two higher-spin conserved tensors implies that
\begin{align}\label{tststb}
[T_{A(s-1),B(s-1)}, T_{C(s-1),D(s-1)}]= &T_{AC}\eta_{AC}...\eta_{BD}...+\mbox{permutations}+\\&+\mbox{other generators possibly}\,,
\end{align}
{\emph{i.e.}} any higher-spin field sources gravity.

The paper is organized as follows. In Section \ref{sec:cubic} we review some known results: the construction of associative HS algebras, non-Abelian deformations of gauge symmetries and the Fradkin-Vasiliev condition. In Section \ref{sec:Jaco} we show that the Jacobi identity is a necessary condition that appears at the quartic level, while its necessity within CFT is obvious. In Section \ref{sec:solvjac} we solve the Jacobi identity with technical details left to the Appendices. As a by-product we classify all non-Abelian parity-violating cubic vertices in $AdS_5$. An extensive discussion of the results is given in Section \ref{sec:conclusions}.

\section{Review of previous results}
\label{sec:cubic}

\subsection{Associative higher-spin algebras}
All known HS algebras, \cite{Konshtein:1988yg, Konstein:1989ij, Eastwood:2002su, Vasiliev:2003ev, Vasiliev:2004cm}, result from certain associative algebras by considering the commutator Lie subalgebra. This is because these algebras are maximal symmetries of linear conformally-invariant equations. The algebra of symmetries of a linear equation is automatically an associative algebra. On the other hand, the best one can prove for the algebra of symmetries of nonlinear equations without any additional assumption is that the symmetry algebra is a Lie algebra (Lie algebroid to be precise). For example, from \cite{Eastwood:2002su}, starting with a linear equation $E(\phi)=0$, a symmetry transformation is an operator $S$ such that it maps solutions to solutions, {\emph{i.e.}} $E\circ S(\phi)=R\circ E(\phi)$ for some other operator $R$. Provided $E$ is a linear operator, one can conclude that given two symmetries $S_1$ and $S_2$, then $S_1\circ S_2$ is a symmetry again. However, the algebra of symmetries is not free: two symmetries $S_1$ and $S_2$ are equivalent if they differ by terms proportional to the linear equations of motion $E(\phi)$, {\emph{i.e.}} $S_1\sim S_2$ if $S_1=S_2+L\circ E$ for some $L$. Therefore, starting with an initial set of symmetries,  conformal symmetries in our case, which are naturally associated with the generators $T_{AB}$ of the conformal algebra, and assuming irreducibility of the system, one may start to multiply them, {\emph{i.e.}} consider products of the type $T_{AB}\circ ... \circ T_{CD}$, generating certain associative algebra. This algebra must then be a quotient of the universal enveloping algebra $U(so(d-1,2))$ by the annihilator of a given module of the conformal algebra that the solution space carries.

The working example of a HS algebra for which a full nonlinear theory is known, \cite{Vasiliev:1990en, Vasiliev:2003ev}, is given by the algebra of symmetries of a free conformal scalar field
\be\square \phi(x)=0\,,\ee
for which one can check \cite{Eastwood:2002su} that there are the following relations among powers of $T_{AB}$
\begin{align}T^{[AB}\circ T^{CD]}&\sim0\,, & C_2=-\frac12 T_{AB}\circ T^{AB}&\sim -\frac{(d-3)(d+1)}4\,,
& T\fud{A}{C}\circ T^{AC}\sim\frac2{d+1}\eta^{AA}C_2&\end{align}
These relations imply that $T_{A(s-1),B(s-1)}=T_{AB}\circ...\circ T_{AB}$ transforms as an irreducible $so(d-1,2)$-tensor with the symmetry of \parbox{42pt}{\boldpic{\RectBRow{4}{4}{$s-1$}{}}}.

An associative HS algebra can always be obtained from $U(so(d-1,2))$
and it is in correspondence with certain linear conformally-invariant equations.
On the other hand, the very existence of a HS algebra that is \textit{essentially} a Lie algebra would correspond to a conformally-invariant \textit{nonlinear} equation whose set of symmetries includes some higher derivative non-linear operators. In this case, the fact that we are searching for an algebra realized on two-row rectangular $so(d-1,2)$-Young diagrams suggests that the corresponding equation is imposed on a scalar field, while the fact the generators are traceless tensors implies that the equation has the following schematic form: $
\square \phi+\mbox{nonlinearities}=0
$.

Let us stress once again, that the associative HS algebra with the required spectrum is essentially unique and
is reproduced by the universal enveloping construction, see e.g. \cite{Iazeolla:2008ix} and references therein. So we are looking for a HS algebra that is a Lie algebra not necessarily originating from $U(so(d-1,2))$.

\subsection{Non-Abelian cubic vertices and $\diamond\,$-products}
\label{star}

In this section we briefly review the results obtained in \cite{Boulanger:2012dx}. Non-Abelian cubic couplings of HS fields are in one-to-one correspondence with possible r.h.s. of commutator $[T_{s_1},T_{s_2}]=...$, where $T_s\equiv T_{A(s-1),B(s-1)}$.
To construct cubic vertices we used the Fradkin-Vasiliev
procedure \cite{Fradkin:1986qy,Fradkin:1987ks} with the manifestly AdS covariant frame-like approach to higher spin fields \cite{Vasiliev:2001wa}. In this framework a massless spin-$s$ field is described by a one-form
$W^{s}$ taking values in the rank $2\check{s}$, $\check{s}:=s-1$,
traceless tensors of $so(d-1,2)$ and
possessing the symmetry of a two-row rectangular Young tableau\footnote{Our convention is that all indices belonging to a group of symmetric (or to be symmetrized) indices are denoted by the same letter. The number of symmetric (to be symmetrized) indices is indicated in brackets. The indices that belong to different groups of symmetric indices are separated by comma.}
\begin{eqnarray}
W^s \leftrightsquigarrow W^{A(s-1),B(s-1)}\;,\qquad \qquad W^{A(s-1),AB(s-2)}=0\qquad
W\fudu{A(s-3)C}{C}{,B(s-1)}=0\,.
\end{eqnarray}
Following \cite{Vasiliev:2003ev}, it is convenient at this point to introduce a set of $2(d+1)$ bosonic oscillators  $Y^A_{\alpha}$,
$\alpha=1,2$ and contract them with tensor indices we deal with. This will enable us
to replace manipulations with tensors by manipulations with corresponding generating
functions. For example, the spin-$s$ field will be encoded by a polynomial
of degree $2\check s$
\begin{equation}
W^{\check s}(Y) :=  \tfrac{1}{(s-1)!(s-1)!}\;
W^{A(s-1),B(s-1)} Y_A^1\ldots Y_A^1\;
Y_B^2\ldots Y_B^2\,.
\end{equation}
The main feature of the $Y$-oscillators is that they can also be used to realize the $\mathfrak{sp}(2)$ algebra \cite{Vasiliev:2003ev},
\begin{equation}
[K_{\alpha\beta}\,,\,K_{\gamma\delta}] =
\epsilon_{\gamma(\alpha}K_{\beta)\delta}+
\epsilon_{\delta(\alpha}K_{\beta)\gamma}\;,
\end{equation}
where
\begin{equation}
K_{\alpha\beta} := \frac{{\rm i}}{2}\;\left( Y^A_\alpha
\frac{\partial}{\partial Y^{\beta\, A}} +
Y^A_\beta
\frac{\partial}{\partial Y^{\alpha\, A}}\right) \;.
\end{equation}
Indices of $sp(2)$ can be raised and lowered by the $sp(2)$-invariant
symbol $\varepsilon^{\alpha\beta}=-\varepsilon^{\beta\alpha}$ according to the rule
$Y^\alpha = \epsilon^{\alpha\beta}Y_\beta\,$,
$Y_\alpha = Y^\beta\epsilon_{\beta\alpha}\,$ where
$\epsilon^{12}=1=\epsilon_{12}\,$. With this notation, the condition that $W^{s}$
is a two-row rectangular Young diagram amounts to:
\begin{equation}
\label{sp2singlet}
[K_{\alpha\beta}\,,\,W^{\check s}(Y)] = 0\,,
\end{equation}
or, in other words, to the statement that higher spin fields are described by $sp(2)$ singlets.

It was shown that the classification of non-Abelian cubic vertices amounts to the classification
of all bilinear products denoted by $\diamond$ that act within a set of traceless tensors with the symmetry of two-row
rectangular Young diagrams:
\begin{equation}
\label{sp2cov}
\parbox{42pt}{\boldpic{\RectBRow{4}{4}{$\check{m}$}{}}}\diamond \parbox{42pt}{\boldpic{\RectBRow{4}{4}{$\check{n}$}{}}}= \bigoplus_k N_{\check{m},\check{n}}^{\check{k}} \, \parbox{42pt}{\boldpic{\RectBRow{4}{4}{$\check{k}$}{}}}\,.
\end{equation}
In other words, having two tensors of shapes
as above, we should study all the independent index contractions that result in rectangular
Young diagrams. Here, $N_{\check{m},\check{n}}^{\check{k}}$ are the multiplicities, which can be greater than one. Resorting to the generating function language, we represent them
by generating functions $f^{\check{m}}(Y)$ and $g^{\check{n}}(Z)$ both being $sp(2)$ singlets.
Then, the elementary contraction is given by
\begin{align*}
\tau^{\alpha\beta}_{YZ}&:=\frac{\partial^2}{\partial Y^A_{\alpha}\partial Z^{\phantom{A}}_{A\beta}}\;,
\end{align*}
which should be followed by setting $Y=Z$ after all derivatives have acted on the corresponding generating functions and then by the application of the projector onto the traceless part, which we ignore for a moment. Other contractions can be obtained as polynomials in $\tau$. To produce a contraction that satisfies (\ref{sp2cov}) all the
$sp(2)$-indices of $\tau$ should be contracted in a $sp(2)$
covariant way.  According to \cite{Boulanger:2012dx}, the linearly independent set of contractions satisfying
(\ref{sp2cov}) is given by the polynomials
\begin{align}
f^{\check{m}}(Y)\diamond g^{\check{n}}(Y) &=
 \left. B_{\check{m},\check{n}}(s_{YZ}, p_{YZ}) f^{\check{m}}(Y) g^{\check{n}}(Z)\right|_{Z=Y},
 \quad B_{\check{m},\check{n}}(s,p)=
 \sum_{\alpha,\beta}b^{\alpha,\beta}_{\check{m},\check{n}}\frac{p^{\alpha}s^\beta}{\alpha!\beta!},
\label{diamond}
\end{align}
of generating elements $s$ and $p$
\begin{eqnarray}
s_{YZ}&:=& \tau^{\alpha\beta}_{YZ}
\epsilon_{\alpha\beta}
\;\equiv\; \frac{\partial^2}{\partial Y^A_{1}\partial Z^{\phantom{A}}_{2A}}-
\frac{\partial^2}{\partial Y^A_{2}\partial Z^{\phantom{A}}_{1A}} \;,
\\
 p_{YZ} &:=& \det{(\tau^{\alpha\beta}_{YZ})}
\; \equiv\; \frac{\partial^2}{\partial Y^A_{1}\partial Z^{\phantom{A}}_{1A}}\frac{\partial^2}{\partial Y^B_{2}\partial Z^{\phantom{B}}_{2B}}-\frac{\partial^2}{\partial Y^A_{1}\partial Z^{\phantom{A}}_{2A}}\frac{\partial^2}{\partial Y^B_{2}\partial Z^{\phantom{B}}_{1B}}\;,
 \label{sandt}
\end{eqnarray}
such that the degrees $\alpha$ and $\beta$ of its terms $p^{\alpha}s^{\beta}$
satisfy
\begin{equation}
\label{basiscond}
\alpha+\beta\leqslant min(\check{m},\check{n}).
\end{equation}
The latter requirement stems from the fact that contractions $p^{\alpha}s^{\beta}$
with $\alpha+\beta>min(\check{m},\check{n})$ prove to be linearly dependent
of contractions $p^{\gamma}s^{\delta}$ with $\gamma+\delta\leqslant min(\check{m},\check{n})$
and $\alpha+2\beta=\gamma+2\delta$. With indices made explicit, one $s$ and one $p$ contractions correspond to
\begin{align}\nonumber
(W^{\check{m}} s W^{\check{n}})^{A(\check{m}+\check{n}-1),B(\check{m}+\check{n}-1)} &= W^{A(\check{m}-1)M,B(\check{m})} W\fud{A(\check{n}),B(\check{n}-1)}{M} -W\fud{A(\check{m}),B(\check{m}-1)}{M}W^{A(\check{n}-1)M,B(\check{n})}\\
(W^{\check{m}} p W^{\check{n}})^{A(\check{m}+\check{n}-2),B(\check{m}+\check{n}-2)}&=(W^{A(\check{m}-1)M,B(\check{m}-1)N}-
W^{A(\check{m}-1)N,B(\check{m}-1)M})
W\fud{A(\check{n}-1)}{M}{}\fud{,B(\check{n}-1)}{N}\label{spcontractions}\;.
\end{align}
The multiplicity $N_{\check{m},\check{n}}^{\check{k}}$ of the tensor product depends on $d$ and for $d\geqslant4$ is given by (without loss of generality we can order spins $\check{m}\leqslant\check{n}\leqslant\check{k}$ ):
\be \label{multipl} N_{\check{m},\check{n}}^{\check{k}}=N_{\check{m},\check{k}}^{\check{n}}=N_{\check{n},\check{m}}^{\check{k}}=\left\{
                                          \begin{array}{ll}
                                            1+[\frac{\check{m}+\check{n}-\check{k}}{2}](1-\delta_{d,4}), & {\check{m}+\check{n}\geqslant\check{k};} \\
                                            0, & \hbox{otherwise.}
                                          \end{array}
                                        \right.
\ee
We recall \cite{Boulanger:2012dx} that the above multiplicity
of $\diamond$-product reproduces the multiplicity of all the possible non-Abelian deformations \cite{Bekaert:2010hp} in Minkowski space.
For $d=5$ one can add certain parity-violating couplings, which are discussed in Section \ref{pvcouplings}.

The structure constants defined by the classification of \cite{Boulanger:2012dx}
involve projectors on rectangular traceless tensors. While these
projectors can be omitted inside the cubic action of \cite{Boulanger:2012dx},
they are of course needed for the explicit computation of the left-hand side of the Jacobi identities
and are responsible for serious technical complications.
The $4d$ case, as we discuss below, can be attacked directly thanks to the isomorphism $so(3,2)\sim sp(4)$ and suitable oscillator variables.
Within the $d$-dimensional setup we solved the Jacobi condition on the space
of traceful tensors
and found a unique solution, where the pure-trace tensors form an ideal.
That we work in the space of traceful tensors is our main simplifying assumption for
the treatment of the Jacobi condition; we comment on this issue in Section \ref{sec:conclusions}.
Actually, in order to put to the test our simplifying assumption, we used Mathematica and performed a computation
with traceless tensors
in a particular case involving spin 4.
We found a unique solution, thereby leading us to
conjecture that our simplifying assumption is not restrictive
in the general spin-$s$ case.

As anticipated, the above frame-like analysis has a counterpart within the ambient metric-like approach of \cite{Joung:2011ww}.
This can be appreciated by restricting the deformed gauge transformations induced by the cubic vertices to the subspace of Killing tensors $E^{\check s}(Y)$  associated with metric-like massless HS fields, with the identification $Y_1=U$, $Y_2=X$. Here, $X$ is the ambient space coordinate while $U$ is an auxiliary ambient variable encoding the tensor indices of the corresponding metric-like fields as
\be
E^{\check s}(X,U)\,=\,\frac{1}{\check s!}\,E^{\check s}_{M_1\ldots M_{\check s}}(X)\,U^{M_1}\cdots U^{M_{\check s}}\,.
\ee
For Killing tensors, satisfying the condition
\be
U\cdot\partial_X\,E^{\check s}(X,U)\,=\,0\,,
\ee
besides
the tangentiality and homogeneity conditions
\be
X\cdot\partial_U\,E^{\check s}(X,U)\,=\,0\,,\qquad
(X\cdot\partial_X-U\cdot\partial_U)\,E^{\check s}(X,U)\,=\,0\,,
\ee
one recovers indeed the general solution
\be
E^{\check s}(X,U)\,=\,\frac{1}{s!}\,E^{\check s}_{N_1,\ldots N_{\check s};M_1\ldots M_{\check s}}\,X^{N_1}\cdots X^{N_{\check s}}\,U^{M_1}\cdots U^{M_{\check s}}\,.
\ee
where the traceless constraints follow from the ambient Fierz system
\be
\square E^{\check s}(X,U)\,=\,0\,,\qquad \partial_U\cdot\partial_X\,E^{\check s}(X,U)\,=\,0\,,\qquad \partial_U\cdot\partial_U\,E^{\check s}(X,U)\,=\,0\,.
\ee
The above Killing tensors, by the ambient construction, are in one-to-one correspondence with the one-form  $W^{\check s}(Y)$ and satisfy the same $sp(2)$-singlet conditions \eqref{sp2singlet}. Here, any tensor contraction in eq.~\eqref{diamond} between gauge parameters defines a (metric-like) Killing tensor algebra and the result of \cite{Boulanger:2012dx} can be interpreted saying that the number of independent non-Abelian couplings is in one-to-one correspondence with the number of independent structure constants for the corresponding Killing tensors.

\subsection{Fradkin-Vasiliev (invariant-normed algebra) condition}
\label{tac}

It is easy to see that a contraction
\begin{equation}
\label{oneterm}
b^{\alpha,\beta}_{\check{n},\check{m}}{p^{\alpha}s^\beta}\,,
\end{equation}
acting on tensors of ranks $2\check{m}$ and $2\check{n}$ produces
a tensor of rank $2\check{k}$ with $\check{k}=\check{n}+\check{m}-2\alpha-\beta$.
The analysis of consistency conditions for non-Abelian cubic vertices (Fradkin-Vasiliev condition)
 shows that a term (\ref{oneterm}) should be accompanied by a term
 \begin{equation}
\label{otherterm}
b^{\alpha',\beta'}_{\check{k},\check{n}}{p^{\alpha'}s^{\beta'}}\,,
\end{equation}
where
$\alpha'=\check{n}-\alpha-\beta$ and  $\beta'=\beta$. It maps tensors of
ranks $2\check{k}$ and $2\check{n}$ to rank $2\check{m}$ tensor.
Moreover, the Fradkin-Vasiliev condition imposes that
\begin{equation}
\label{FVnew}
\frac{a(k)}{(\alpha+\beta+1)}\;\frac{b^{\alpha,\beta}_{\check{n},\check{m}}}{\alpha!\beta!}\;=\;
\frac{a(m)}{(\alpha'+\beta'+1)}\;\frac{b^{\alpha',\beta'}_{\check{k},\check{n}}}{\alpha'!\beta'!}\;,
\end{equation}
where $a(k)$ and $a(m)$ are normalization constants that one can always introduce in front of
the quadratic parts of the actions for spin $k$ and spin $m$ fields. The quadratic actions are of the form $\int R\wedge R\wedge...$ with $R$ being the linearized field-strength $R=DW$.
This condition can be thought of as an invariant norm\footnote{The algebras of interest turn out to have a trace, which is stronger than having a bilinear form.} condition for $\diamond$
algebra, \cite{Boulanger:2012dx}.

We will not need its precise form in the following. What will be essential is just the fact that nonzero $b^{\alpha,\beta}_{\check{n},\check{m}}$ implies
nonzero $b^{\alpha',\beta'}_{\check{k},\check{n}}$. Let us also note,
that due to an ambiguity mentioned above, the same term
(\ref{otherterm}) can have different appearances. For the Fradkin-Vasiliev
condition to be satisfied it is enough that (\ref{otherterm})
is present in any of its forms, not necessarily satisfying (\ref{basiscond}).

Roughly speaking, the Fradkin-Vasiliev condition implies that the quadratic on-shell action can be represented as trace $\int \langle C,  C\rangle$, where the norm $\langle,\rangle$ is with respect to the HS algebra and $C$ contains higher-spin Weyl tensors. Then one can prove that the action remains gauge-invariant at the cubic level, {\emph{i.e.}} when the linearized $R=DW$ is replaced with \eqref{fvdeform} and the gauge transformations are properly deformed.

\section{Jacobi identity}
\label{sec:Jaco}

In this section we recall some basic facts about consistent deformations of a free
gauge theory, and refer to \cite{Berends:1984rq,Barnich:1993vg,Boulanger:2000rq}
for more details. In particular, we want to recall that the
associativity of the infinitesimal gauge transformations implies the Jacobi
identity for the gauge algebra, which appears at the second order in
deformation.
The context of this section concerns a generic gauge system with
open algebra\footnote{Note the early works \cite{Fradkin:1977wv,Sterman:1977ds,Kallosh:1978de}
relevant for the concept of open gauge algebra.}
for which the Batalin-Vilkovisky formalism \cite{Batalin:1981jr}
is particularly useful. However, we refrain from resorting to the antifield formalism here
and refer to the book and review \cite{Henneaux:1992ig,Barnich:2000zw}
for information and references.

\paragraph*{Perturbative deformations.}

Using De Witt's condensed notation whereby summation over indices also implies
integration over spacetime, one considers a gauge-invariant action
$S[\{\varphi^i\};g]$ for a set of gauge fields $\{\varphi^i\}$ that propagate and
interact in a given maximally symmetric background with metric components
$\bar{g}_{\mu\nu}\,$,
and where $g$ denotes a (set of) deformation parameter(s), such that in the limit
$g\rightarrow 0\,$ the action $S[\{\varphi^i\};g]$ smoothly reduces to a
positive sum of quadratic actions $S_{0}^i[\varphi^i]\,$,
one for each field $\varphi^i\,$:
\begin{eqnarray}
\lim_{g\rightarrow 0} S[\{\varphi^i\};g] = \sum_{i} S_{0}^i[\varphi^i]\;.
\end{eqnarray}
The action $S[\varphi^i;g]$ is invariant, $\delta_{\epsilon}S=0\,$, under
the gauge transformations
$\delta_\epsilon\varphi^i = R^i_{\alpha}\epsilon^{\alpha}\,$, and the gauge
algebra reads
\begin{eqnarray}
 &\Big( \delta_{\epsilon_2}\delta_{\epsilon_1}
 - \delta_{\epsilon_1}\delta_{\epsilon_2} \Big) \varphi^i
 = 2\,\Big( R^i_{\alpha}\, f^{\alpha}{}_{\beta\gamma}
 + \frac{\delta S}{\delta \varphi^j}\,M^{ji}_{\beta\gamma} \Big)\,
 \epsilon_1^{\gamma}\epsilon_2^{\beta} \;, \quad \mbox{where} &
 \label{gaugealgebra} \\
 &f^{\alpha}{}_{\beta\gamma}  =
 - (-1)^{\mbox{deg}(\beta)\mbox{deg}(\gamma)}  f^{\alpha}{}_{\gamma\beta}\;,
 \quad
 M^{ji}_{\beta\gamma} = -(-1)^{\mbox{deg}(i)\mbox{deg}(j)}M^{ij}_{\beta\gamma}
 =- (-1)^{\mbox{deg}(\beta)\mbox{deg}(\gamma)}M^{ji}_{\gamma\beta}\;.\qquad &
 \nonumber
 \end{eqnarray}
Expanding the left-hand side of Eq. (\ref{gaugealgebra})
and using the fact that the gauge parameters
are arbitrary functions, one obtains
\begin{eqnarray}
 R^j_{\beta}\,\frac{\delta R^i_{\alpha}}{\delta\varphi^j}
 - (-1)^{\mbox{deg}(\beta)\mbox{deg}(\alpha)}
  R^j_{\alpha}\,\frac{\delta R^i_{\beta}}{\delta\varphi^j}
 \,  &=&
 2\,R^i_{\gamma} f^{\gamma}{}_{\beta\alpha}
 +2\,\frac{\delta S}{\delta \varphi^j}\,M^{ji}_{\beta\alpha} \;.
\label{gaugestructure}
\end{eqnarray}
Expanding all the relevant quantities in powers of the deformation parameter(s)
$g\,$ and taking into account the fact that the free action is characterized by an Abelian gauge
algebra,
\begin{eqnarray}
\delta_{\epsilon}\varphi^i &=&  \delta^{(0)}_{\epsilon}\varphi^i  +
g\, \delta^{(1)}_{\epsilon}\varphi^i + g^2\, \delta^{(2)}_{\epsilon}\varphi^i +\ldots
\;=\; \Big( {R}^{(0)\,i}{}_{\alpha} + g\, {R}^{(1)\,i}{}_{\alpha}
+ g^2\, {R}^{(2)\,i}{}_{\alpha} + \ldots \Big) \,\epsilon^{\alpha}\;,
\\
f^{\gamma}{}_{\beta\alpha} &=&  0 + g\, f^{(1) \gamma}{}_{\beta\alpha} +
g^2\, f^{(2) \gamma}{}_{\beta\alpha} + \ldots\;,
\end{eqnarray}
one obtains
\begin{eqnarray}
 R^{(0)\,j}{}_{\beta}\,\frac{\delta R^{(1)\,i}{}_{\alpha}}{\delta\varphi^j}
 - (-1)^{\mbox{deg}(\beta)\mbox{deg}(\alpha)}
 R^{(0)\,j}{}_{\alpha}\,\frac{\delta R^{(1)\,i}{}_{\beta}}{\delta\varphi^j}
 \,  &=&
 2\, R^{(0)\,i}{}_{\gamma} f^{(1) \gamma}{}_{\beta\alpha} \;.
 \label{starone}
\end{eqnarray}

In \cite{Boulanger:2012dx}, the quantities
$ f^{(1) \gamma}{}_{\beta\alpha}$ for totally symmetric higher-spin gauge fields
in the frame-like and manifestly $AdS_d$-covariant formalism
were classified, and part of the quantities
$ R^{(1)\,i}{}_{\alpha}$ that are responsible for the non-Abelian gauge
algebra deformation $ f^{(1) \gamma}{}_{\beta\alpha}$ were also given therein (see \cite{JoungTaronnaToAppear} for the metric-like analysis).

At the next order in the deformation parameter $g\,$, the closure of the
gauge algebra (\ref{gaugestructure}) gives
\begin{eqnarray}
 R^{(1)\,j}{}_{[\beta}\, (\delta R^{(1)\,i}{}_{\alpha]} / \delta\varphi^j) \,  +
 R^{(0)\,j}{}_{[\beta}\, \delta R^{(2)\,i}{}_{\alpha]} / \delta\varphi^j \,
 &=&
  R^{(1)\,i}{}_{\gamma} \,f^{(1) \gamma}{}_{\beta\alpha} +
    R^{(0)\,i}{}_{\gamma} \,f^{(2) \gamma}{}_{\beta\alpha}
 +\, \frac{\delta S_0^j}{\delta \varphi^j}\,M^{(0) ji}_{\;\;\beta\alpha} \;.
\nonumber
\end{eqnarray}
Taking the linearized gauge transformation $\delta^{(0)}$ of this equation,
performing the complete (graded) antisymmetry over the free indices
corresponding to the three gauge parameters and using (\ref{starone}),
one derives
\begin{eqnarray}
R^{(0)\,j}{}_{\delta} \,f^{(1) \delta}{}_{[\gamma\beta}
(\delta R^{(1)\,i}{}_{\alpha]} / \delta\varphi^j) &=&
R^{(0)\,i}{}_{\mu} \,f^{(1) \mu}{}_{[\gamma\vert \delta}
f^{(1) \delta}{}_{\vert\beta\alpha]}
+
R^{(0)\,i}{}_{\delta} \,R^{(0)\,j}{}_{[\gamma} \,
 (\delta f^{(2) \delta}{}_{\beta\alpha]} / \delta\varphi^j)\;,
\end{eqnarray}
where the terms proportional to $R^{(2)\,i}{}_{\delta}$ drop out
because $R^{(0)\,(j}{}_{[\beta} \,R^{(0)\,k)}{}_{\gamma]}\equiv 0\, $.
Using Eq. (\ref{starone}) again on the left-hand side of
the above equation, one obtains
\begin{eqnarray}
2\, R^{(0)\,i}{}_{\sigma} \,f^{(1) \sigma}{}_{\delta [\alpha}
f^{(1) \delta}{}_{\gamma\beta]} &=&
R^{(0)\,i}{}_{\sigma} R^{(0)\,j}{}_{[\gamma}
(\delta f^{(2) \sigma}{}_{\beta\alpha]} / \delta\varphi^j)\;.
\end{eqnarray}
Discarding an irrelevant constant term, this yields the Jacobi condition we
were looking for:
\begin{eqnarray}
f^{(1) \sigma}{}_{\delta [\alpha}
f^{(1) \delta}{}_{\gamma\beta]} &=&
\tfrac{1}{2}\,R^{(0)\,j}{}_{[\gamma}
(\delta f^{(2) \sigma}{}_{\beta\alpha]} / \delta\varphi^j)\;.
\label{JacobiFirst}
\end{eqnarray}
We note that this equation can be derived more elegantly using the cohomological
reformulation \cite{Barnich:1993vg} of the consistent deformation procedure
and was used in the higher-spin context in \cite{Bekaert:2005jf,Boulanger:2005br,Bekaert:2006us,Boulanger:2006gr,Bekaert:2010hp}.
The presence of a non-vanishing right-hand side
is a common feature of higher-spin gauge fields in the metric-like formulation and
is easily recognized as a linearised gauge transformation.

An advantage of the frame-like formalism
used in \cite{Boulanger:2012dx}, namely, the MacDowell-Mansouri-Stelle-West-Vasiliev formalism, is that the right-hand side of
the above equation is zero, since the quantities
$f^{(1) \sigma}{}_{\delta \alpha}$ are purely algebraic
(they do not act as differential operators),
whereas the operators $R^{(0)\,j}{}_{\alpha}$ do act as a differential.
This is to be contrasted with the metric-like formulation of higher-spin fields
where the equivalent quantities  $f^{(1) \gamma}{}_{\alpha\beta}$ act as differential
operators, see e.g. \cite{Bekaert:2005jf,Boulanger:2005br,Boulanger:2006gr,Bekaert:2010hp,Taronna:2011kt}.

In the metric-like formalism, it is nevertheless possible to draw a link with the frame-like one and look at a simplified problem by restricting the attention to the subspace of Killing tensors of the free theory, as described above.
In such a case, the right-hand side of (\ref{JacobiFirst}) vanishes.
[The point
with the Mac-Dowell-Mansouri-Stelle-West-Vasiliev frame-like
formalism is that the
equation  (\ref{JacobiFirst}) has a vanishing right-hand side,
{\emph{without}} imposing any restriction on the space of gauge parameters on which it applies.]
The resulting metric-like conditions can then be rephrased as the
``formalism-independent'' requirement that the rigid-symmetry algebra
has to be a Lie algebra.

In the following, we will impose the condition (\ref{JacobiFirst}) as a
restriction
on all the possible ${f^{(1) \gamma}}_{\alpha\beta}$'s that we have
classified in \cite{Boulanger:2012dx}, and see which of them pass the Jacobi-identity test.


\section{Solving the Jacobi identity}
\label{sec:solvjac}

In this Section we solve the Jacobi identity for a $\diamond$-commutator
\begin{equation}
\label{intrcom}
[f^{\check{m}},g^{\check{n}}]_{\diamond}=f^{\check{m}}\diamond g^{\check{n}}-
g^{\check{n}}\diamond f^{\check{m}}= C_{\check{m},\check{n}}(s_{YZ}, p_{YZ}) f^{\check{m}}(Y) g^{\check{n}}(Z)|_{Y=Z}.
\end{equation}
In what follows, equating the arguments after the contractions have been performed, like
$Y=Z$ in the above formula, will be implicit for brevity. We will also use
a power series decomposition
\begin{equation}
\label{powerserfors}
 C_{\check{n},\check{k}}(s,p)=\sum_{i,j}c^{j,i}_{\check{n},\check{k}}\frac{s^i}{i!}
 \frac{p^j}{j!}\,.
\end{equation}

First, as a warm-up exercise we will solve the problem with the simplifying assumption
that the structure function $C$ does not depend on the spins of the
fields it acts upon.
The general case is more technical. It is presented in the Appendix A.

The basic contractions $s$ and $p$ introduced previously have the following symmetry
properties under the exchange of arguments:
\begin{equation}
\label{symps}
s_{YZ}=-s_{ZY}\,, \qquad \qquad p_{YZ}=p_{ZY}\,.
\end{equation}
The partial derivatives are distributive under replacing one of the factors with a
product of two other factors, which gives
\begin{equation}
f(X)\tau_{XY}g(Y)h(Y)=f(X)(\tau_{XY}+\tau_{XZ})g(Y)h(Z)\,.
\end{equation}
This entails distributivity properties for $s$ and $p$
\begin{equation}
f(X)s_{XY}g(Y)h(Y)=f(X)(s_{XY}+s_{XZ})g(Y)h(Z)\,,
\end{equation}
\begin{equation}
f(X)p_{XY}g(Y)h(Y)=f(X)(p_{XY}+p_{XZ}+\tau_{XY}\cdot \tau_{XZ})g(Y)h(Z)\,,
\end{equation}
where we find a new  $sp(2)$-invariant contraction involving
three  tensors $\tau_{XY}\cdot \tau_{XZ}
:=\tau^{\alpha\beta}_{XY}\cdot \tau^{\gamma\delta}_{XZ}\varepsilon_{\beta\gamma}\varepsilon_{\alpha\delta}$.
As a consequence, there is a normal distributivity for $s$, as it is linear in each of the
derivatives. The distributivity for $p$ is however violated by an extra term.

The antisymmetry of a commutator together with (\ref{symps}) implies
\begin{equation}
C(s,p)=-C(-s,p)\,.
\end{equation}
Once $C$ is known, the $s$-odd part of $B$ in \eqref{diamond}
is also fixed
\begin{equation}
\label{candb}
b^{i,j}_{\check{m},\check{n}}=\tfrac{1}{2}c^{i,j}_{\check{m},\check{n}}\,, \quad j \quad
 \text{odd}\,,
\end{equation}
 while the $s$-even
part remains undetermined.

The Jacobi identity then reads
\begin{align}
\notag
\Big[C(s_{XY}+s_{XZ},p_{XY}+p_{XZ}+\tau_{XY}\cdot \tau_{XZ})C(s_{YZ},p_{YZ})&-\\
\label{jacoperform}
C(s_{YZ}-s_{XY},p_{YZ}+p_{XY}+\tau_{YZ}\cdot \tau_{YX})C(s_{XZ},p_{XZ})&-\\
\notag
C(s_{XZ}+s_{YZ},p_{XZ}+p_{YZ}+\tau_{XZ}\cdot \tau_{YZ})C(s_{XY},p_{XY})&\Big]
f(X)g(Y)h(Z)=0\,.
\end{align}
We observe immediately that each of the three types of linearly independent $\tau\cdot \tau$
contractions appears only in one term and thereby, if present, cannot be cancelled.
So, the only way to satisfy Jacobi identity is when $C(s,p)$ is $p$-independent, see however comments in Section \ref{sec:conclusions}.
Then, denoting $\{s_{XY},s_{ZX},s_{YZ}\}$ as $\{x,y,z\}$ for brevity, and omitting
$f$, $g$ and $h$, we obtain
\begin{equation}
\label{jacoperform1}
C(x-y)C(z)+C(z-x)C(y)+C(y-z)C(x)=0\,.
\end{equation}
To solve it we first act with $\partial_z$ and put $z=0$ afterwards getting
\begin{equation}
\label{jacoperform2}
C(x-y)C'(0)=-C'(x)C(y)+C(x)C'(y)\,.
\end{equation}
Assuming that $C'(0)=0\,$, we find that $C'(x)/C(x)=const=\alpha$, which implies
$C(x)=\gamma e^{\alpha x}$, and is incompatible with the original assumption.
So, $C'(0)\ne 0$. Acting on (\ref{jacoperform2}) with $\partial_y$ and setting $y=0$ we find
\begin{equation}
C(x)C''(0)=0 \quad \Rightarrow \quad C''(0)=0\,.
\end{equation}
On the other hand, by acting on (\ref{jacoperform2}) with $\partial^2_y$
and setting $y=0$ we obtain
\begin{equation}
C''(x)/C(x)=C'''(0)/C'(0)\,.
\end{equation}
There are two cases: $C'''(0)=0$ and $C'''(0)\ne 0$. Taking into
account that $C(x)$ is odd, the general solution reads
\begin{align}
\label{poissonsol}
C(x) =&\gamma x\,, &    \text{if}& & C'''(x)&=0\,,\\
\label{moyalsol}
C(x) =&\gamma \,\text{sinh}(\alpha x)\,,  &    \text{if}& & C'''(x)&\ne 0\,.
\end{align}
The second solution is the Lie algebra associated to the Moyal product. The
first one is Poisson bracket, that results from the Moyal commutator in the
limit $\alpha\rightarrow 0$ with $\alpha\gamma$ fixed.

Let us note that the Poisson
algebra does not satisfy the Fradkin-Vasiliev condition in $d\geqslant4$, which was mentioned already in \cite{Fradkin:1987ks}. Indeed,
the $\diamond$-commutator contains
$c^{0,1}_{\check{m},1}\ne 0$. Non-zero value of this coefficient is responsible
for the correct transformation properties of the spin $m$ field under $so(d-1,2)$. It then
implies that the $\diamond$ product contains non-zero $b^{0,1}_{\check{m},1}$.
Then, the Fradkin-Vasiliev condition requires that
either $b^{\check{n}-1,1}_{\check{n},\check{n}}$ or one of its forms should
be non-zero. However, all the coefficients $c^{\check{n}-1-a,1+2a}_{\check{n},\check{n}}$
are zero for the Poisson solution. Hence the consistency condition cannot be satisfied.
On the contrary, for the Moyal product $b^{0,2\check{n}-1}_{\check{n},\check{n}}\ne 0$,
and the Fradkin-Vasiliev condition is fulfilled. Roughly-speaking, the Poisson bracket contracts one pair of indices and cannot reproduce \eqref{tststb} that requires $2s-1$ pairs of indices to be contracted.

Note that, in the context of deformation quantization, the uniqueness of the Moyal bracket on phase-space is a fairly well-known result, so that
if one asks that  $C(s,p)$ start with $C(s,p)=\alpha s + \ldots $ and
satisfy the Jacobi condition, then the unique result, up to ordering freedom is the Moyal bracket, see e.g.
\cite{Bayen:1977pr,Bayen:1977ha,Arverson1983,Fletcher:1990ib,Cohen1966,Cohen1976,ZFC} where we recommend the beautiful book \cite{ZFC} for extensive
references and very pedagogical exposition of quantum  mechanics on
phase space.

\paragraph{Trace condition.} Let us stress here that the latter analysis has been carried out on the space
of traceful tensors, therefore leaving aside possible solutions that are intrinsically defined on traceless
tensors and for which no extension to traceful tensors is available. The corresponding analysis on
traceless tensors is much more involved due to the complicated structure of traceless projectors.
This type of computation can be carried out with the help of the xAct and xTras packages for Mathematica \cite{xAct1,xAct2,xAct3,Nutma:2013zea} and we have explicitly solved numerically the spin-4 components of the Jacobi identity ending up with a unique solution that is in correspondence with the traceful one.
The result obtained with the help of Mathematica
therefore confirms the uniqueness of the Eastwood-Vasiliev agebra
on traceless tensors.
A summary of the Mathematica computation can be found in Appendix B.

We remark that our results on the uniqueness of the simplest Vasiliev algebra
can be interpreted in various ways, depending on which coupling we assume to be nonzero to start with
and which coupling is therefore deduced from the Jacobi condition.
One could start by saying that the explicit computation does
not rely on any assumption about the gravitational coupling \eqref{transformastensors}
and consequently on the presence in the HS algebra of the
$so(d-1,2)$ subalgebra, and instead assumes that there is
a nontrivial spin-4 self-coupling.
In other words, the Jacobi identity for spin 4 implies that once one of the
structure constants is non-zero,  then all the others are fixed uniquely.
We conjecture the same property for a general setup. It would be
interesting to clarify it elsewhere.

\paragraph{Dimension dependent identities.}
Till now the analysis has been carried out independently of the dimension of $AdS_d$. However, depending on the dimension $d$ it is well known that the appearance of Schouten identities of the form\footnote{Without loss of generality one can concentrate on singlet identities, which are generated by contracting all the indices of \eqref{DDI} in order to get a scalar quantity. In \eqref{JacobiFirst} one can reinstall the three gauge parameters $\epsilon^\alpha, \eta^\beta, \xi^\gamma$ and further contract with $k_{\sigma\lambda}\zeta^\lambda$, for $k_{\sigma\lambda}$ the components of the invariant metric. This is possible since, both at the cubic and at the quartic level using the invariant norm, one can deal only with scalar quantities. Moreover, any other tensor identity can be recovered from the scalar ones using the properties of the tensor product.}
\be\label{DDI}
I_d:\quad \delta^{[a_1}_{b_1}\cdots\delta^{a_{d+1}]}_{b_{d+1}}\Big|_d\equiv0\,,
\ee
might produce further sporadic solutions. In the following we analyze all possible cases in which the above identities may or are known to play a role when solving the Jacobi identity. We shall see that our analysis is valid and directly applicable only in $d=4$ and $d>6$.

\subsection{Two- and three-dimensional cases}
Our arguments cannot be easily applied to two- and three-dimensional cases due to the appearance of the fundamental Schouten identities\footnote{Notice that the vanishing of the Weyl tensor as well as all similar d=2,3 identities can be generated from them.} $I_2$ and $I_3$, \eqref{DDI}.
The latter can play a role both at the level of cubic couplings and at the level of the Jacobi identity. For instance, it is well known that for $d\leqslant 3$ two derivative couplings involving higher-spin fields do exist, radically modifying the cubic coupling classification itself\footnote{For instance, among the other things, one can prove that the Schouten bracket, that is generically inconsistent in $d\geqslant 4$, satisfies the Fradkin-Vasiliev conditions in $d=3$. See e.g. ref \cite{Boulanger:2005br} where Schouten identities played a central
role in the context of cubic vertices and Jacobi identities for HS fields in flat background.}. In these cases, however, key simplifications come from the fact that one can automatically take into account such Schouten identities by noticing that the corresponding AdS algebras can be realized as $so(1,2)\sim sp(2)$ and as $so(2,2)\sim sp(2)\oplus sp(2)$, respectively. Using the latter isomorphisms, one can then find one-parameter families of algebras $hs(\nu)$ and $hs(\nu)\oplus hs(\nu)$ \cite{Prokushkin:1998bq, Vasiliev:1989re} that are defined as quotients of the universal enveloping algebra of $sp(2)$ by the two-sided ideal generated by the Casimir operator \cite{Feigin}:
\be
hs(\nu)=U(sp(2))/(C_2-\nu)\,.
\ee

\subsection{Four-dimensional case}

In $d=4$, two-derivative couplings involving higher-spins cease to exist but the list of vertices is much shorter then in higher dimensions due to the fundamental Schouten identity $I_4$, \eqref{DDI}.
As in $d=2,3$, it is however more convenient to automatically take care of the above identity exploiting spinorial
representations and taking advantage of the isomorphism $so(3,2)\sim sp(4)$. Indeed, the irreducible representations of
$so(3,2)$ defined by two-row rectangular Young diagrams turn out to be equivalent to totally-symmetric tensors of $sp(4)$\,:
\begin{align}
so(3,2)&:\parbox{42pt}{\boldpic{\RectBRow{4}{4}{$k$}{}}} && \Longleftrightarrow & sp(4)&: \parbox{42pt}{{\RectARow{4}{$2k$}{}}}\,.
\end{align}
Notice also that symmetric tensors of $sp(4)$ are automatically irreducible, since the $sp(4)$ metric tensor $C^{\Lambda\Omega}$, $\Lambda=1,..,4$, is antisymmetric. Therefore, by working within the $sp(4)$ language, we actually automatically restrict the algebra to traceless tensors without the need of considering projectors.

More concretely, the tensor product in the class of totally-symmetric tensors of $sp(4)$ is multiplicity free, \eqref{multipl}, {\emph{i.e.}} given three spins $s_1$, $s_2$, $s_3$ there can be at most one non-Abelian vertex. Back to the $so(3,2)$ language, the $p$-contraction disappears as an independent object due to eq.~\eqref{DDI}, making clear why there is only one non-Abelian vertex possible for given three spins.

In order to work with $sp(4)$, instead of the auxiliary doublet $Y^A_\alpha$, we need a singlet $Y^\Lambda$ which is a vector of $sp(4)$,
\be
W(Y)=\frac1{(2s-2)!} W^{\Lambda(2s-2)} Y_\Lambda\ldots Y_\Lambda\,.
\ee
Hence, one can see that there is only one elementary contraction
\be
\tau_{YZ}=\frac{\pl}{\pl Y^\Lambda}C^{\Lambda\Omega}\frac{\pl}{\pl Z^\Omega}\,,
\ee
that obeys the same properties as the $s$-contraction, {\emph{i.e.}} it is distributive. In
vectorial notations $\tau_{YZ}$ is the same as the $s$ contraction of Section \ref{star}.
Therefore, when solving the Jacobi identity, one can proceed as in the case of arbitrary $d$ with the result that the $4d$ HS algebra is
unique and comes from the associative Moyal-Weyl $\star$-product,
\be
f(Y)\star g(Y) = \left .f(Y) \exp{\left(\alpha \tau_{YZ}\right)} g(Z)\right|_{Z=Y}\label{fdalgebra}\,.
\ee
This algebra underlies the Vasiliev $4d$ theory \cite{Vasiliev:1990en} and was originally found in \cite{Fradkin:1986ka} solving directly the Jacobi identity using the Lorentz-covariant basis of $sl(2,\mathbb{C})\subset sp(4)$.

Let us comment on the degeneracy that occurs in $CFT_3$. While in dimension greater than three the spectra of the single trace operators in bosonic and fermionic free vector models are different, \cite{Vasiliev:2004cm}, they almost coincide in three dimensions. The OPE's $\phi\times\phi$ and $\psi\times\psi$ contain conserved tensors of all (even) ranks. The only difference is in the scalar operator $j_0$ that has weight one and two for the bosonic and fermionic vector models, respectively. In addition, the Casimir operators of the free conformal scalar and fermion are the same, $C_2=-5/4$. Therefore, the universal enveloping construction yields identical HS algebras in both cases, which can be realized as the algebra of even functions of $Y^\Lambda$ under the product \eqref{fdalgebra}. All that being said, the sole $4d$-HS algebra admits two fundamental representations: free scalar and free fermion, which was observed at the level of correlation functions in \cite{Maldacena:2011jn}.

\subsection{Five-dimensional case}

In $5d$ the Schouten identity $I_5$, \eqref{DDI}
does play a role at the cubic level allowing for a parity-violating vertices that we discuss below and
it does play a role for the Jacobi identity that is a quartic object for which the number of different groups of symmetrized indices is 7.
In $d=5$ however one can benefit from the isomorphism between $so(4,2)$
and $su(2,2)$ and implement automatically the above identities using the spinor oscillators $b^{\alpha}$ and $a_{\alpha}$,
transforming in the fundamental and the conjugated fundamental
representations of $su(2,2)$, respectively,
\cite{Fradkin:1989xt,Sezgin:2001zs,Vasiliev:2001zy,Vasiliev:2001wa}.
It proves that an $so(4,2)$ irreducible tensor of shape \parbox{42pt}{\boldpic{\RectBRow{4}{4}{$s-1$}{}}}
can be represented by an $su(2,2)$-tensor
with $s-1$ vector and $s-1$ covector indices:
\begin{equation}
\label{spinyd}
W^{A(s-1),B(s-1)} \quad \rightarrow \quad
W^{\alpha(s-1);}{}_{\beta (s-1)}\,, \qquad\qquad W^{\alpha(s-2)\gamma;}{}_{\beta (s-2)\gamma}\equiv0
\end{equation}
which is symmetric in each set of indices. In addition it must be traceless. The necessity for the trace constraint complicates the $5d$ story. Fortunately, the trace projector in $su(2,2)$ basis is the $sp(2)$-extremal projector, which is simpler than the $sp(4)$-extremal projector needed for general $d$.

It is natural to combine higher spin fields into generating functions as
\begin{equation}
W^{\check s}(a,b) :=  \tfrac{1}{(s-1)!(s-1)!}\;
W^{\alpha(s-1);}{}_{\beta (s-1)} a_\alpha\ldots a_\alpha\;
b^\beta\ldots b^\beta\,.
\end{equation}

Starting with two generating functions  $f^{\check{m}}(a,b)$ and $g^{\check{n}}(c,d)$
one can construct two basic contractions
\begin{equation}
\label{uv}
u_{(ab)(cd)}:= \frac{\partial^2}{\partial a_\alpha \partial c^{\alpha}}\,, \qquad
v_{(ab)(cd)}:= \frac{\partial^2}{\partial b^\beta \partial d_{\beta}}\,,
\end{equation}
that both close in the space of higher spin generators, {\emph{i.e.}}
the result of their action is a sum of terms each being of the form (\ref{spinyd}), while the following combinations
\begin{equation}
\label{st}
s_{(ab)(cd)}:=u_{(ab)(cd)}-v_{(ab)(cd)}\,, \qquad t_{(ab)(cd)}:=u_{(ab)(cd)}+v_{(ab)(cd)}\,,
\end{equation}
have the following symmetry properties under the exchange of arguments
\begin{equation}
\label{exch}
s_{(ab)(cd)}=-s_{(cd)(ab)}\,, \qquad t_{(ab)(cd)}=t_{(cd)(ab)}\,.
\end{equation}
The appearance of the $t$-contraction corresponds in vectorial $so(6)$-notation to the following basic contraction
\be
W^{A(s_1-2)U,B(s_1-2)U} \epsilon\fdu{UUVV}{AB} W^{A(s_2-2)V,B(s_2-2)V}\label{tcontraction}\,,
\ee
which is a 'square root' of the $p$-contraction. Notice that in contrast to the $p$-contraction, the $t$-contraction obeys the distributivity property, hence simplifying the structure of the jacobiator \eqref{jacoperform}.

Finally, the general expression for a $\diamond$-commutator in spinorial terms is
 \begin{equation}
\label{intrcomsp}
[f^{\check{m}},g^{\check{n}}]_{\diamond}=f^{\check{m}}\diamond g^{\check{n}}-
g^{\check{n}}\diamond f^{\check{m}}= C_{\check{m},\check{n}}(s_{(ab)(cd)}, t_{(ab)(cd)})
f^{\check{m}}(a,b) g^{\check{n}}(c,d)|_{(ab)=(cd)},
\end{equation}
with
\begin{equation}
\label{antisymcomps}
 C_{\check{m},\check{n}}(s,t)=-C_{\check{n},\check{m}}(-s,t)\,.
\end{equation}
which follows from the antisymmetry of the commutator and from (\ref{exch}). Eventually, one can show
that $s$ and $p$ contraction of the vectorial approach map to $s$ and
$i(s^2 - t^2)$ contractions in spinorial language, respectively.

Our $d$-dimensional considerations cannot be applied directly to the $d=5$ case because $p$ gets replaced with $t$, due to the identity of eq.~\eqref{DDI}. In other words the $\tau\cdot\tau$ contractions of eq.~\eqref{jacoperform} are not any more independent. We comment further on this issue in the Conclusions, recalling the existence of a one-parameter family of HS algebras in $AdS_5$.

\subsection{Six-dimensional case}
The $d=6$ case is the highest dimensional case in which Schouten identity $I_6$, \eqref{DDI}, may play a role for symmetric tensors. As in the $5d$ case, $I_6$ does not influence the classification of cubic couplings. However, it might play some role at the level of the Jacobi identity \eqref{jacoperform}, since at the quartic order we have exactly $7$ different groups of symmetrized indices. In this case, differently of before, there is no isomorphism that might help in carrying out the classification and we just state that our analysis does not apply directly. We leave for the future a more detailed study of this case in which, however, no additional solution to the one that we have found above is known.

In dimensions higher than $d=6$ all Schouten identities require the anti-symmetrizations of more than $7$ indices. Hence, no non-trivial identity can appear for symmetric tensors even at the level of Jacobi identity. To summarize, we have seen that our analysis can be directly applied to the $d=4$ and $d>6$ cases.

\subsection{Parity violating HS algebras}\label{pvcouplings}

Depending on the dimension of $AdS_d$, in parallel to Schouten identities, one might also consider parity-violating cubic couplings that can be constructed with the help of the totally-antisymmetric tensor $\epsilon_{a_1\ldots a_d}$ (odd number thereof). We do not consider the very special 3-dimensional case, so that $d \geqslant 4\,$.
A similar reasoning to the one applied to Schouten identities tells us that such parity-violating coupling can only arise in $d \leqslant 5$. The classification of parity-violating couplings has been carried out in flat space in \cite{Bengtsson:1986kh, Metsaev:2005ar,Boulanger:2005br}, however no corresponding classification in AdS is available. In principle the presence of non-Abelian parity-violating couplings might be associated to corresponding parity violating HS algebras that however are outside the scope of the present paper. Let us just mention few more details about the $d=4,5$ cases.

\paragraph{Four-dimensions.} It is well known that at cubic level there is a one parameter family of Vasiliev's theories each of which is conjectured to be dual to a corresponding $CFT_3$ where the parity breaking is realized through a Chern-Simons term, \cite{Giombi:2011kc}. What we have proved above\footnote{There are two types of parity-violating vertices in $4d$. One is 'Weyl tensor cubed' and another one is low-derivative coupling that should be non-Abelian. The latter coupling is seen in the light-cone approach while it does not seem to have any Lorenz-covariant analog. We are grateful to R.Metsaev for the important comments on this issue. }, exploiting the spinor language, is that all non-Abelian vertices are parity-preserving and hence all the members of the parity-violating class of higher-spin theories are based on the same HS algebra while the one-parameter family (infinite family if one takes into account also higher orders) arises from Abelian parity-breaking couplings that are available in 4d.

\paragraph{Five dimensions.} It is straightforward to generalize \cite{Vasilev:2011xf, Boulanger:2012dx} so as to include parity-violating non-Abelian vertices. There are two elementary contractions in $5d$, the usual $s$-contraction, \eqref{spcontractions}, \eqref{st}, and the $t$-contraction, \eqref{tcontraction}, the latter involves the $\epsilon$-symbol. Obviously, any even power of the $t$ contraction results in a parity-preserving vertex. In order to make a parity-odd singlet out of three tensors having the symmetry of rectangular two-row Young diagrams we have first to contract six indices with the $\epsilon$ symbol, which takes away two indices from each of the tensors or one column from each of the Young diagrams. Therefore, given three spins $s_1$, $s_2$, $s_3$ the number of non-Abelian parity-violating vertices in $5d$ is given by the number of non-Abelian parity-preserving vertices among fields with spins $s_1-1$, $s_2-1$, $s_3-1$. We see that there is enough room for parity-violating HS algebras, {\emph{i.e.}}
HS algebras that are $\mathbb{Z}_2$ graded where the grade-zero component corresponds to a parity-preserving HS algebra that is realized on the grade-one subspace while the bracket on the grade-one piece is nontrivial.

\section{Conclusions and Discussion}\label{sec:conclusions}

When pushing the analysis of cubic vertices to the next, quartic order,
one finds that the gauge algebra with
(graded)-antisymmetric structure constants should obey the Jacobi identity. The conserved charges in CFT have also to obey the Jacobi identity. The latter is automatically satisfied if the commutator arises from an underlying
associative structure.
It is not mandatory that the gauge algebra should arise
in that way, though.
In the present note, we show that this is however the case.

This can be argued, but not proved, as follows. Given a higher-spin theory whose spectrum contains massless AdS fields, assuming AdS/CFT, one can find a dual CFT with conserved
higher-spin currents. By the axioms of CFTs, the conserved higher-spin currents should appear in the OPE $\phi\times\phi$ of some field, say $\phi$, that in order to ensure the current conservation condition should obey a linear equation of the form
\be\label{linear eq}
\square \phi+\ldots=0\,.
\ee
Representation theory tells that $\phi\otimes \phi$ decomposes into infinitely many conserved higher-spin tensors, hence the AdS dual must contain higher-spin gauge fields with arbitrary large spins. Here, the fact that $\phi$ obeys a linear equation is the counterpart of the fact that the symmetry algebra, which is generated by higher-spin charges, is an associative algebra. As we mentioned in the introduction this does not imply the triviality of the bulk theory.
Indeed, the HS algebra gets deformed at the nonlinear level and as
a Lie algebra it is not necessarily a symmetry of the HS theory.

In \cite{Vasiliev:2011xf, Boulanger:2012dx} the
non-Abelian vertices for symmetric higher spin fields in $AdS$ space were classified.
It was shown in \cite{Boulanger:2012dx} that each of them is
associated with a $\diamond$-product
that acts within the space of two-row
rectangular $so(d-1,2)$ Young diagrams.
In the present paper we explicitly solved the Jacobi identity
for the commutator associated with the $\diamond$-product showing that
there is only one $\diamond$-product
that passes this consistency test --- the one constructed by
means of the Moyal-Weyl star-product. Thereby,
the only cubic vertex that has a chance to be promoted
to the next order is the one associated with the so-called
``$s$-contraction"
rule of Section \ref{star}, where the latter
contraction rule is the germ for
the associative algebra used by Vasiliev in constructing
the full nonlinear HS theory in \cite{Vasiliev:2003ev}.
Proving the uniqueness of the algebraic structure that underlies
higher spin interactions, we provide a strong evidence
 of uniqueness of the full theory proposed in
\cite{Vasiliev:1990en,Vasiliev:1992av,Vasiliev:2003ev}.

The advantage of having a theorem is that it becomes more clear which of the assumptions should be relaxed in order to find new solutions. Our results indicate that if there exists some other HS algebra in addition to the Eastwood-Vasiliev one, then it must be due to one of the following reasons.

With the help of the trace/Killing norm all the problems can be reduced to singlets. In particular, taking the trace of the candidate commutator $[b,c]$ with an auxiliary generator $a$ we get a trilinear form
\be \YoungpAAA: \qquad C(a,b,c)=Tr(a[b,c])\,,\ee
which is totally antisymmetric in the three slots. Analogously, the Jacobi identity is equivalent to vanishing of
\be \YoungpAAAA: \qquad C(a,b,c,d)=Tr([a,b][c,d]+\mbox{three more})\,,\ee
which is totally antisymmetric in the four slots. While the possible structures that can contribute to the commutator can be easily classified, there are nontrivial identities (different from the Schouten identities) mentioned in \cite{Vasiliev:2011xf} that could produce further sporadic solutions. Therefore, new solutions, if any, should result from combining $p$-contractions with the nontrivial identities at the quartic level. However, we do not expect them to play a role for the Jacobi constraint since those identities should involve pure quartic traces of the type $Tr(abcd)$ that by construction do not appear in Jacobi due to its factorized structure.

Lower dimensions up to $6d$ escape our arguments due to Schouten-like identities. Exceptionally, $4d$ case is under control thanks to the isomorphism $sp(4)\sim so(3,2)$ that trivializes $p$-contractions. The $4d$ result is quite strong since it is about on-shell algebras, {\emph{i.e.}}
 the algebras realized on irreducible traceless generators. In higher-dimensions, where there are no other relations but those that are mentioned in \cite{Vasiliev:2011xf},  our main result is about algebras on traceful tensors.

In more details, the most general $\diamond$-product can be constructed
not only in terms of $s$ and $p$ basic contractions, but
can also contain various terms that take and produce traces.
For the cubic vertex construction these terms were not important
because the initial diagrams were traceless and the result of
$\diamond$-multiplication was projected to the traceless
part by contraction with traceless tensors. Working with
the $\diamond$-product on traceful tensors, we
have shown that there is only one solution whose commutator satisfies the Jacobi identity.
The resulting Lie algebra admits traces as an ideal, so by quotienting them
we end up with a Lie algebra that acts on traceless tensors. But it is still not clear whether all Lie algebras acting on traceless tensors can be constructed in this way. There might exist such a Lie algebra that essentially acts on traceless tensors and \emph{cannot} result from a quotienting of a traceful one.
Our explicit computation shows that the sector of spin-four is protected and produces no new solutions as compared to the case of traceful tensors. We leave this interesting issue for further research.

To summarize, the main gaps are concerned with the diversity of $5d$ (including parity-violating structures), which we discuss below, $6d$ and the issue of trace projections in Jacobi identity for spins higher than four\footnote{It would be more than surprising to find that there is a nontrivial solution starting from spin six or one hundred.}.

\paragraph{Zoo of HS algebras.} Let us recall that in $2d$ and $3d$ one has a one-parameter family of HS algebras $hs(\nu)$ and $hs(\nu)\oplus hs(\nu)$ that come from associative algebras, via quotients of $U(sl(2))$.

Curiously enough there exist finite-dimensional HS algebras \cite{Boulanger:2011se}, that obey all our assumptions. In particular, these algebras lead to theories that are consistent at least to the cubic level. In general, they contain generators corresponding to mixed-symmetry fields with the only exception of $5d$, where the spectrum truncates to totally-symmetric fields. A $5d$ example of such an algebra, which corresponds to $hs_5(9)$ of \cite{Boulanger:2011se}, was considered recently in \cite{Manvelyan:2013oua}. The known finite-dimensional HS algebras exists for odd $d$ only,
{\emph{i.e.}} for $so(d+1)$ belonging to $D_n$ series, and are given by universal enveloping algebras of the self-dual representation with weights $(k,...,\pm k)$.

Relaxing the assumptions about the spectrum of generators and allowing partially-massless fields one finds a zoo of finite-dimensional algebras given by universal enveloping algebra of representations with weights $(k,0,...,0)$, {\emph{i.e.}} totally symmetric rank-$k$ traceless tensors.

If we are not confined to the class of totally-symmetric fields allowing for mixed-symmetry fields, but still insist on unitarity, then, the universal enveloping algebra $U(so(d-1,2))$ provides a one-parameter family of algebras $hs_d(\nu)$, \cite{Boulanger:2011se}. The unitarity assumption restricts the Young symmetry types of the generators as gauging some of them may lead to partially-massless or nonunitary mixed-symmetry fields, \cite{Alkalaev:2003qv,Skvortsov:2006at,Skvortsov:2009zu}, and results in the family $hs_d(\nu)$. When $d=5$ or $d=7$ analogous algebras from a different perspective were obtained in \cite{Fernando:2009fq, Fernando:2010dp}. This family includes the infinite-dimensional symmetry algebras of various conformally-invariant equations, classified in \cite{Shaynkman:2004vu}, which is analogous to the Eastwood-Vasiliev algebra resulting from a conformal scalar. In this respect, finite-dimensional HS algebras can be viewed as symmetries (endomorphisms) of finite-dimensional representations of the conformal-algebra, which can be defined as solutions to certain conformally-invariant equations which are over-determined and have a finite-dimensional space of solutions. While there are strong indications that one cannot have HS theories with a finite number of fields in the spectrum
(in $d>3$ where HS fields become propagating), the finite-dimensional solutions mentioned above should be taken into account since they formally solve the Jacobi identity.

All the known algebras with mixed-symmetry fields in the spectrum share the property that totally-symmetric higher-spin fields are always present. It is expected that our methods can be straightforwardly generalized to the case of mixed-symmetry fields with the same conclusion about the necessity of totally-symmetric higher-spin fields in the theories with mixed-symmetry fields.

\paragraph{Further constraints on HS algebras.}  There is a more accurate test for HS algebras whereby finite-dimensional algebras are expected to fail: the admissibility condition, \cite{Konshtein:1988yg, Konstein:1989ij}.

The admissibility condition requires the candidate HS algebra to have a (unitary) representation that under the $so(d-1,2)$ subalgebra decomposes precisely into the same set of particles as is given by the gauging of this algebra. The admissibility condition is naturally implemented within the unfolded approach, \cite{Vasiliev:1988xc, Vasiliev:1988sa}, which is at the core of the Vasiliev HS theory
and, as well, within the Noether deformation procedure
where it comes as the next condition after the Jacobi identity.
The deformation procedure within the unfolded approach implies that HS gauge connections
belong to a certain Lie algebra, while zero-forms (gauge-invariant on-shell field-strengths) form a module of this algebra. The space of zero-forms is dual to the space of one-particle states, \cite{Shaynkman:2001ip, Vasiliev:2004cm}. Therefore, the admissibility condition is just one of the structure equations that constrains possible deformations within the unfolded approach. As a problem of representation theory
\cite{Konshtein:1988yg, Konstein:1989ij,Vasiliev:2004cm}
the admissibility condition
can be studied more effectively than the consistency of the Lagrangian up to the quartic level.

In particular, all finite-dimensional HS algebras come together with the fundamental representation, which is finite-dimensional too. As such it (and all its tensor powers) fails to pass the admissibility test. In addition, one may argue, that HS theories with boundary conditions preserving full amount of HS symmetry are dual to free CFT's and free CFT's have conserved tensors of arbitrary high rank, which seems in contradiction with the finite number of HS fields in the bulk.

\paragraph{Fate of other cubic vertices.} Interesting is also the fate of the other Abelian and non-Abelian cubic vertices. First of all, let us reiterate that only the non-Abelian vertices turn out to be relevant for the problem of classifying HS algebras while the Jacobi identity is only one of the several conditions to be satisfied at the quartic level. Certain combinations of Abelian cubic vertices needs to be added in order to make the quartic vertex consistent, while others might just parameterize families of inequivalent theories and this indeed happens in the $4d$-Vasiliev theory, which has infinitely many free parameters (one for the cubic and quartic levels), if the parity symmetry is sacrificed \cite{Vasiliev:1990en}. The latter are associated to Abelian parity violating couplings. Such couplings do not exist at the cubic level in $d>5$. The $d$-dimensional Vasiliev theory, \cite{Vasiliev:2003ev}, as it is constructed, does not seem to have any ambiguity different from field redefinitions. However, for any given dimension there exists an order at which one may attempt to construct a parity-violating vertex. It is still possible that one can deform the $d$-dimensional Vasiliev theory by some parity-violating Abelian interactions.

Concerning the rest of non-Abelian vertices, they all seem to be left forbidden: there is no room for the other types of non-Abelian vertices in the theory of totally-symmetric HS fields. We expect that these non-Abelian vertices should appear in theories containing mixed-symmetry fields in their spectrum, where the classification of non-Abelian vertices needs to be reconsidered. See, however, \cite{Boulanger:2012dx} for a conjecture. Indeed, the $p$-contraction ($p$ to the first power, to be precise) does contribute to the product of the HS algebra of \cite{Vasiliev:2004cm} that is defined as the symmetry algebra of a free conformal fermion in $(d-1)$ dimensions. As we mentioned, it is in $CFT_3$ only that the symmetries of a free boson and free fermion coincide, while in higher dimensions the OPE's are essentially different, with $\psi\times \psi$ OPE containing certain mixed-symmetry tensors as well \cite{Vasiliev:2004cm}. The HS algebra of \cite{Vasiliev:2004cm} again results from the universal enveloping construction. The lesson is that non-associative $\diamond$-products admit an associative resolution provided the spectrum of generators is enlarged. We beleive that all powers of $p$-contraction can be successively resolved with higher-spin singletons, see, e.g. \cite{Bekaert:2009fg}.

\paragraph{Correlation functions.} Finally let us mention that given a HS algebra one can immediately write down an answer for the $n$-point correlation functions \cite{Colombo:2012jx, Didenko:2012tv}
\begin{align} \langle j ... j\rangle&=Tr(\Phi\circ ...\circ \Phi), & \delta \Phi&=[\Phi,\xi] \label{corrfunc}\end{align}
where $\Phi$ transforms in the adjoint of the HS algebra and is related to the boundary-to-bulk propagator of the corresponding HS theory. The correlator \eqref{corrfunc} is totally fixed by the HS symmetry and does not require the knowledge of the full HS theory in AdS, see \cite{Colombo:2012jx, Didenko:2012tv, Gelfond:2013xt, Didenko:2013bj} for examples of computations, which are much simpler than the perturbative computations in Vasiliev theory \cite{Giombi:2009wh, Giombi:2010vg}.

\section*{Acknowledgements}

We thank X. Bekaert, M. G\"{u}naydin, E. Joung, R. Metsaev, K. Siampos, Ph. Spindel and P. Sundell for useful discussions. We are grateful to D. Fairlie and C. Zachos for very kind explanations
and correspondence on the issue of the uniqueness of Moyal's bracket. We are indebted to M.A. Vasiliev for useful discussions and comments on the manuscript. We thank Teake Nutma for his help with xAct and xTras. The work of N.B. and D.P. was supported in part by
an ARC contract No. AUWB-2010-10/15-UMONS-1. The work of E.S. was
supported  in part by the Alexander von Humboldt Foundation, by RFBR grant
No.11-02-00814, 12-02-31837. We thank the Galileo Galilei Institute for Theoretical Physics for the hospitality and the INFN for partial support during the completion of this work.

\appendix
\renewcommand{\thesection}{Appendix \Alph{section}}
\renewcommand{\theequation}{\Alph{section}.\arabic{equation}}
\setcounter{equation}{0}\setcounter{section}{0}

\section{Solving the Jacobi identity}
In this Appendix we solve the Jacobi identity for the $\diamond$ commutator
with spin-dependent coefficients (recall that the function $C$ in (\ref{intrcom})
may depend on $\check{m}$ and $\check{n}$). Due to the same arguments as
in Section \ref{sec:solvjac} one can derive, that the coefficient functions
should be $p$-independent. Then, for brevity, let us introduce the notation
$c^i_{\check{n},\check{k}}=c^{0,i}_{\check{n},\check{k}}$,  see (\ref{powerserfors}).
The Jacobi identity reads:
 \begin{equation}
     \label{tosolve}
\text{Jac}(\check{m},\check{n},\check{k})=\sum_i{C_{m,n+k-i}(x-y)c^i_{n,k}\frac{z^i}{i!}+
 C_{n,k+m-i}(z-x)c^i_{k,m}\frac{y^i}{i!}+
 C_{k,m+n-i}(y-z)c^i_{m,n}\frac{x^i}{i!}=0}\,,
 \end{equation}
 where, as before,
  \begin{equation}
 \label{parity}
 C_{\check{m},\check{n}}(s)=-C_{\check{n},\check{m}}(-s)\,.
 \end{equation}
Our aim is to show that these equations imply
the existence of a
scale redefinition of the algebra generators
such that the coefficient functions
do not depend on the spin. After that we can apply the solution given in
Section \ref{sec:solvjac}.

Being more general compared to (\ref{jacoperform}), eq. (\ref{tosolve}) contains
additional solutions to those found for spin-independent coefficients.
For example, it admits a Virasoro-like solution
\begin{equation*}
C_{\check{m},\check{n}}=\alpha (\check{m}-\check{n})\,,
\end{equation*}
where $\alpha$ is a constant. Various truncations of Poisson algebra, such as a truncation that contains only
spin $2$ and spin $3$ generators, also solve (\ref{tosolve}).
However, the first class of solutions does not contain $so(d-1,2)$ as a subalgebra,
while the second class of solutions does not satisfy the Fradkin--Vasiliev condition.

Our aim is to show that,
%
assuming the existence of at least one higher spin $m>2$ field that couples
minimally to gravity --- see \eqref{transformastensors} and footnote \ref{footnote7} ---,
so that
 \begin{equation}
 \label{orighs}
 c^0_{1,\check{m}}=0\,, \quad c^1_{1,\check{m}}\ne 0\,, \quad c^2_{1,\check{m}}=0\,,
 \end{equation}
together with the Fradkin-Vasiliev condition that we use as
 \begin{equation}
 \label{FV}
 c^1_{1,\check{m}}\ne 0
 \quad   \Leftrightarrow   \quad  c^{2\check{m}-1}_{\check{m},\check{m}}\ne 0\,,
 \end{equation}

\noindent one can deduce that:
\begin{itemize}
\item in case I, when the initial higher spin field is of odd spin, then all other
fields are present in the spectrum and the only solution of
(\ref{tosolve}) is (\ref{moyalsol});
\item in case II, when the initial higher spin field is of even spin, then
one finds also a solution (\ref{moyalsol}) of (\ref{tosolve}),
 but truncated to even spins.
\end{itemize}

In what follows, we will show that above requirements lead to
\begin{itemize}
\item In case I
\begin{equation}
 \label{2}
 c^0_{\check{k},\check{l}}=0\,, \quad \forall \ \check{k},\check{l}
 \end{equation}
and
\begin{equation}
\label{conj2}
c^1_{\check{m},2}\ne 0\,, \quad \forall\ \check{m}\,.
\end{equation}
The latter condition implies that there is a non-zero contraction, that acts
on spin $m$ and spin $3$ fields producing spin $m+1$ field. So, in essence
it implies that all the spins are present in the spectrum. By making relative
rescaling of spin $m$ and spin $m+1$ generators one can always set
\begin{equation}
\label{3}
c^1_{\check{m},2}=1.
\end{equation}
Differentiating (\ref{tosolve}) with respect to
$z$ and setting $y=z=0$ we then find
 \begin{equation}
 \label{y0z1}
 C_{\check{m},\check{n}+\check{k}-1}(x)c^1_{\check{n},\check{k}}+
 C'_{\check{n},\check{k}+\check{m}}(-x)c^0_{\check{k},\check{m}}
 -\sum_i c^1_{\check{m}+\check{n}-i,k}c^i_{\check{m},\check{n}}\frac{x^i}{i!}=0\,.
 \end{equation}
 From (\ref{y0z1}), for $\check{k}=2$ and (\ref{2}), (\ref{3}) one finally recovers
 \begin{equation}
 \label{4}
 C_{\check{m},\check{n}+1}(x)=C_{\check{m},\check{n}}(x)\,,
 \end{equation}
 which was to be proved.

\item In case II
\begin{equation}
 \label{2prime}
 c^0_{\check{k},\check{l}}=0\,, \quad \forall \check{k},\check{l} \quad \text{odd}
 \end{equation}
and
\begin{equation}
\label{conj3}
c^1_{\check{m},3}\ne 0\,, \quad \forall \check{m} \quad \text{odd}\,.
\end{equation}
The latter condition implies that there is a non-zero contraction, that acts
on spin $m$ and spin $4$ fields producing a spin $m+2$ field. So,
it implies that all the even spins are present in the spectrum. By making relative
rescalings of spin $m$ and $m+2$ generators one can always set
\begin{equation}
\label{5}
c^1_{\check{m},3}=1\,.
\end{equation}
Hence, from (\ref{y0z1}) for $k=3$ and (\ref{2prime}), (\ref{5}) one finds
 \begin{equation}
 \label{6}
 C_{\check{m},\check{n}+2}(x)=C_{\check{m},\check{n}}(x)\,,
 \end{equation}
 which was to be proved.

\end{itemize}

\paragraph*{Final steps.}
To complete the proof we need to show that (\ref{orighs}), (\ref{FV}),
together with Jacobi identity (\ref{tosolve}) imply either (\ref{2}), (\ref{conj2})
or (\ref{2prime}), (\ref{conj3}), depending on parity of the initial spin that is assumed to be present.

 Eq. (\ref{tosolve}) can be decomposed into power series of $x$, $y$ and $z$.
 The equation appearing as a coefficient in front of $x^py^qz^r$ reads
 \begin{equation}
 \label{tosolve1}
 c^{p+q}_{\check{m},\check{n}+\check{k}-r}c^r_{\check{n},\check{k}}(-1)^q+
 c^{r+p}_{\check{n},\check{k}+\check{m}-q}c^q_{\check{k},\check{m}}(-1)^p+
 c^{q+r}_{\check{k},\check{m}+\check{n}-p}c^p_{\check{m},\check{n}}(-1)^r=0\,.
 \end{equation}
 We should also keep in mind, that these operators act on finite Young diagrams.
 The conditions that the above equation is not satisfied trivially, due to the fact that all the contractions
 require more indices than it is available, read as
 \begin{equation}
 \label{val}
 q+r\leqslant 2\check{k}, \quad p+q\leqslant 2\check{m}, \quad p+r\leqslant 2\check{n}\,.
 \end{equation}

 Suppose $\check{m}$ is in the spectrum, let us study
  $\text{Jac}(\check m,\check m,\check m)=0$.
  One can then choose $\check m=\check n=\check k$ in
 (\ref{tosolve1}), which gives
  \begin{equation}
 \label{st1}
 c^{p+q}_{\check{m},2\check m-r}c^r_{\check m,\check m}(-1)^q+
 c^{r+p}_{\check m,2\check m-q}c^q_{\check m,\check m}(-1)^p+
 c^{q+r}_{\check m,2\check m-p}c^p_{\check m,\check m}(-1)^r=0\,.
 \end{equation}
 First, we check the above equation for $p=2\check m-1$, $q=1$, $r=0$. In this case one gets
   \begin{equation}
 \label{st2}
 c^{2\check m}_{\check m,2\check m}c^0_{\check m,\check m}(-1)+c^{2\check m-1}_{\check m,2\check m-1}c^1_{\check m,\check m}(-1)+
 c^{1}_{\check m,1}c^{2\check m-1}_{\check m,\check m}=0\,.
 \end{equation}
 Taking into account that $c^0_{\check m,\check m}=0$ from (\ref{parity}),
 $c^{1}_{\check m,1}\ne 0$ from (\ref{orighs}) and $c^{2\check m-1}_{\check m,\check m}\ne 0$
 from (\ref{FV}) one finds from (\ref{st2}) that
 \begin{equation}
 \label{st3}
 c^1_{\check m,\check m}\ne 0\,,\quad c^{2\check m-1}_{\check m,2\check m-1}\ne 0\,.
 \end{equation}

 Let us now set $r=1$, $p=2i$, $q=2\check m-2i-1$, where $i<\check m$ in (\ref{st1}):
   \begin{equation}
 \label{st4}
 c^{2\check m-1}_{\check m,2\check m-1}c^1_{\check m,\check m}(-1)^q+c^{2i+1}_{\check m,2i+1}c^{2\check m-2i-1}_{\check m,\check m}+
 c^{2\check m-2i}_{\check m,2\check m-2i}c^{2i}_{\check m,\check m}(-1)=0\,.
 \end{equation}
 The first term is nonzero due to (\ref{st3}), the last term is zero because
 $c^{2i}_{\check m,\check m}=0$. This allows to find that
 \begin{equation}
 \label{st5}
 c^{1+2i}_{\check m,1+2i}\ne 0\,, \quad c^{2\check m-2i-1}_{\check m,\check m}\ne 0\, \quad \forall\ i\,, \quad 0\leqslant i<\check m\,.
 \end{equation}
 Each of the inequalities in (\ref{st5}) allows to say that once $f^{\check m}$
 is present in the spectrum, then all other fields $g^{\check n}$ with an odd
 $\check n$
 in a range from $0$ to $2\check m$ are present too. Applying this argument iteratively,
 one can see that if one higher spin is present in the spectrum, then
 all even spins are present in the spectrum too.

 Taking in (\ref{st1}) $p=2s$, $q=2i$, $r=2j+1$, one can find that:
 \begin{equation*}
 c^{2(i+s)}_{\check m,2\check m-2j-1}c^{2j+1}_{\check m,\check m}=0\,,
 \end{equation*}
 which implies
 \begin{equation}
 \label{st6}
 c^{2(i+s)}_{\check m,2\check m-2j-1}=0\,.
 \end{equation}
 Here, $2\check m-2j-1$ is any odd number in the range $0, \dots, 2\check m$, while $2(i+s)$
 is any even number.

 From now on the proof will split for case I and case II.

 \paragraph*{Case II}
In this case the initial $\check{m}$ is odd.
Then, as it was just shown,
 all other odd generators are present. Eq. (\ref{st6}) then implies that all
 $c$ with even upper index vanish.
 In particular, (\ref{2}) holds.

 Let us now consider $\text{Jac}(\check m,\check m,\check n)=0$ for odd $\check m$ and
 $\check n$
   \begin{equation}
 \label{st7}
 c^{p+q}_{\check m,\check m+\check n-r}c^r_{\check m,\check n}(-1)^q
 +c^{r+p}_{\check m,\check n+\check m-q}c^q_{\check n,\check m}(-1)^p+
 c^{q+r}_{\check n,2\check m-p}c^p_{\check m,\check m}(-1)^r=0\,.
 \end{equation}
 Next, we put $p,q,r$ so as
 \begin{equation}
 \label{st07}
 2\check m=p+q+r\,, \quad r=1\,, \quad \text{$q$ is even and $p$ is odd}\,.
 \end{equation}
 In this case, the second term in (\ref{st7}) vanishes because $c^q_{\check n,\check m}=0$
  (\ref{st6}),
 while the last term is nonzero, due to (\ref{st5}). So, the first term is nonzero,
 which entails
 \begin{equation}
 \label{st007}
 c^1_{\check m,\check n}\ne 0\,, \quad \forall \quad \check m\,,\check n\,.
 \end{equation}
 From the latter one finally gets eq.~(\ref{conj3}). Hence, one can apply the logic above and see that the
 $C_{\check m,\check n}$ are independent from $\check m$ and $\check n$.

 \paragraph*{Case I}

 Let us now consider the case when the original $\check m$ is even.
 As we have already shown, all odd generators are present in the spectrum.
 Now, first, we will show that all the even generators are present as well.
 Then we will show that coefficients $c^{2i}_{\check{k},\check{l}}$
 are zero for all $\check{k}$ and $\check{l}$. Finally, we will show an analog
 of (\ref{st007}) where $\check m$ and $\check n$ are not supposed to be odd any more.

 Let the original $\check m$ be $2l$. Then $2l-1$
 is also in the spectrum and we look at $\text{Jac}(2l,2l,2l-1)=0$:
  \begin{equation}
 \label{st8}
 c^{p+q}_{2l,4l-1-r}c^r_{2l,2l-1}(-1)^q+c^{r+p}_{2l,4l-1-q}c^q_{2l-1,2l}(-1)^p+
 c^{q+r}_{2l-1,4l-p}c^p_{2l,2l}(-1)^r=0\,.
 \end{equation}
 For $p$ odd $c^p_{2l,2l}$ is non-zero (\ref{st5}). For $c^{q+r}_{2l-1,4l-p}$ to be non-zero
 we set
 \begin{equation}
 \label{st9}
 q+r=4l-p\,,
 \end{equation}
 see (\ref{st5}), which also requires
 \begin{equation}
 \label{st09}
 4l-p\leqslant 2(2l-1)-1=4l-3\,, \quad \text{so} \quad p\geqslant 3\,.
 \end{equation}
 Since $4l-p$ is odd, $q$ and $r$ should be of opposite parity. Let us say that
 $q$ is even, $r$ is odd. Then, the second term in (\ref{st8}) vanishes because
 $c^q_{2l-1,2l}=0$ for even $q$, (\ref{st6}). So we find
 \begin{equation}
 \label{st10}
 c^r_{2l,2l-1}\ne 0\,,
 \end{equation}
  \begin{equation}
 \label{st11}
 c^{p+q}_{2l,4l-1-r}\ne 0\,,
 \end{equation}
 which implies, that $4l-1-r$ belongs to the spectrum. Recall that $r$ satisfies (\ref{st9}),
 where $p\geqslant 3$ (\ref{st09}) and $q$ is a positive even integer. This implies that $1\leqslant r\leqslant 4l-3$ and
 consequently $2\leqslant 4l-1-r\leqslant 4l-2$. So, we have shown, that from the fact that $2l$
 is present in the spectrum follows the presence of all even generators from $2$ to $4l-2$.
 Applying this argument iteratively one finds that all even generators are in the spectrum.

 Taking $p=2s$, $q=2i$, $r=2j+1$ in (\ref{st8}) one finds
 \begin{equation}
 \label{st12}
 c^{2(l+s)}_{2l,4l-2-2j}=0\,,
 \end{equation}
 which means that even power contractions between any even generators are vanishing.

 Let us denote by $e_i$ and $o_j$ indices that take only even and odd values correspondingly.
In this terms,  up to now we have shown that
 \begin{equation}
 \label{sum1}
 c^{e_1}_{o_1,o_2}=0\,, \quad \forall \quad o_1, o_2, e_1\,,
 \end{equation}
 which follows from (\ref{st6}) for $\check m$ odd, and
  \begin{equation}
 \label{sum2}
 c^{e_1}_{e_2,e_3}=0\,, \quad \forall \quad e_1, e_2, e_3\,,
 \end{equation}
 which follows from (\ref{st12}). One can now use (\ref{st6}) for $\check m$ even,
 ariving at
  \begin{equation}
 \label{sum3}
 c^{e_1}_{e_2,o_1}=0\,, \quad \forall \quad o_1, e_1, e_2:
 \quad  o_1 \leqslant 2 e_2-1\,.
 \end{equation}
 What is left is to eliminate the last constraint between $o_1$ and $e_2$.
  To this end we consider $\text{Jac}(2l-1,2l,2l-1)=0$:
   \begin{equation}
 \label{st13}
 c^{p+q}_{2l-1,4l-1-r}c^r_{2l,2l-1}(-1)^q+c^{r+p}_{2l,4l-2-q}c^q_{2l-1,2l-1}(-1)^p+
 c^{q+r}_{2l-1,4l-1-p}c^p_{2l-1,2l}(-1)^r=0.
 \end{equation}
 As usual, we set $p=2s$, $q=2i$, $r=2j+1$ and obtain
  \begin{equation}
 \label{st14}
 c^{2(l+s)}_{2l-1,4l-2-2j}=0\,,
 \end{equation}
 which implies
   \begin{equation}
 \label{sum4}
 c^{e_1}_{e_2,o_1}=0\,, \quad \forall \quad o_1, e_1, e_2:
 \quad  e_2 \leqslant 2 o_1\,.
 \end{equation}
 Together with (\ref{sum3}) the above equation covers all the possible relative values
 of $e_2$ and $o_1$.

 To sum up,
    \begin{equation}
 \label{sum5}
 c^{e_1}_{\check m,\check n}=0\,, \quad \forall \quad e_1, \check m, \check n\,,
 \end{equation}
 and in particular, (\ref{2}) is true.

Now, we use (\ref{st7}), where $\check m$ and $\check n$
 are not supposed to be odd any more. Starting from
(\ref{st07}) we find (\ref{st007}) and, in particular, (\ref{conj2}).
So, all the assumptions of the derivation  (\ref{2}), (\ref{conj2})  have
been proven and we conclude, that (\ref{4}) is also true.

\section{Analysis on traceless tensors}

In order to check the absence of further solutions to the Jacobi identity that might be intrinsically defined
on traceless tensors we have concentrated for simplicity on the equation $\text{Jac}[3,3,3]=0$
corresponding to external spin-4 totally-symmetric higher-spin particles. The latter equation has been
projected on its spin-4 component involving only a finite number of higher-spin structure constants
${f^{(1)\alpha}}_{\gamma\delta}$:
\be
{f^{(1)4}}_{2[4}{f^{(1)2}}_{44]}+{f^{(1)4}}_{4[4}{f^{(1)4}}_{44]}+{f^{(1)4}}_{6[4}{f^{(1)6}}_{44]}=0\,,
\ee
where we are restricting the attention for simplicity to even-spins since we know that with such
restriction we can already find solutions.

In this way it has been possible to restrict the number of independent tensor structures contributing to
the Jacobi identity still recovering a non-trivial system of equations for the relative coefficients of the
most general Ansatz for the structure constants. In order to proceed, we have worked from the very
beginning on the space of traceless two-rows Young tableaux using explicit traceless projectors on the
various tensor components and parameterizing the most general structure constants without further
assumptions. We have then found one tensor structure for ${f^{(1)2}}_{44}\sim\, s^5$ that we have
parameterized with a coefficient $\alpha_1$, two tensor structures for
${f^{(1)4}}_{44}\sim \, s^3,\, s\, p$
that we have parameterized with two coefficients\footnote{The basis that we have chosen
on the space of traceless tensors is different from the basis given in terms of the building blocks
$s$ and $p$.}
$\gamma_1$ and $\gamma_2$, and finally a single tensor structure for ${f^{(1)6}}_{44}\sim\, s$
that we have parameterized with a constant $\epsilon_1$.
The latter is the counterpart of the classification \cite{Boulanger:2012dx}
of non-Abelian cubic couplings in terms of the basic building blocks $s$ and $p\,$.
Starting from the structure constants we have computed separately the three contributions to the Jacobi
identity above corresponding to spin-2, spin-4, and spin-6. After summing all contributions we have got
a sum of $1560$ independent tensor contractions whose coefficients are set independently to zero by
the Jacobi identity. The latter requirement fixes the couplings constants in terms of $\gamma_1$ as:
\begin{eqnarray}
&&\alpha_1=\pm\frac{\sqrt{d(d-2)}}{2
\sqrt{(d+1)(d+3)}}\frac{(d+5)}{5d+16}\,\gamma_1\,,\\
&&\gamma_2=-\frac{2(d-4)}{5d+16}\,\gamma_1\,,\\
&&\epsilon_1=\pm\frac{5\sqrt{(d+2)(d+4)(d+5)}}{6\sqrt{d+7}(5d+16)}\,\gamma_1\,,
\end{eqnarray}
where the sign ambiguity is a consequence of the quadratic nature of the particular component of the
Jacobi identity we have analyzed and reflects also some phase ambiguity that is present in the Moyal
solution found above.
As a consequence, assuming a spin-4 field generator ($\gamma_1\neq 0$) automatically forces gravity,
as well as a spin-6 generators into the spectrum with a precise choice of the relative coefficients for their
non-Abelian couplings (structure constants). The latter matches the analogous result that we have
found on traceful tensors leading us to conjecture
that no further solution exists, that is defined
intrinsically on traceless tensors.

\providecommand{\href}[2]{#2}\begingroup\raggedright\endgroup


\end{document}